%% file: main.tex
\documentclass[sigconf]{acmart}
\usepackage{color}
\usepackage{listings}
\usepackage{amsmath,amsfonts}
\usepackage[ruled, vlined, linesnumbered]{algorithm2e}
\SetAlFnt{\small}
\SetAlCapFnt{\small}
\SetAlCapNameFnt{\small}
\usepackage{multirow}
\usepackage{makecell}
\usepackage{subcaption}
\usepackage{enumitem}
\usepackage{xcolor}
\definecolor{CodeBG}{RGB}{255,255,255}
\definecolor{CodeFrame}{RGB}{255,255,255}
\definecolor{CodeNums}{RGB}{125,135,156}
\definecolor{CodeKW}{RGB}{0,161,124}
\definecolor{CodeType}{RGB}{0,91,142}
\definecolor{CodeStr}{RGB}{38,92,66}
\definecolor{CodeCmt}{RGB}{0,142,255}
\definecolor{CodeEmph}{RGB}{120,0,0}

\lstdefinestyle{acmcode}{
  backgroundcolor=\color{CodeBG},
  frame=single, rulecolor=\color{CodeFrame}, framerule=0.4pt,
  xleftmargin=1.8em, framexleftmargin=0.8em, framesep=0.6em,
  numbers=left, numberstyle=\footnotesize\color{CodeNums}, numbersep=6pt,
  basicstyle=\ttfamily\scriptsize,
  keywordstyle=\bfseries\color{CodeKW},
  commentstyle=\itshape\color{CodeCmt},
  stringstyle=\color{CodeStr},
  emphstyle=\bfseries\color{CodeEmph},
  tabsize=2, keepspaces=true, columns=fullflexible,
  showstringspaces=false, breaklines=true, breakatwhitespace=true,
  upquote=true,
  escapeinside={||}{||} % keep your |...| escapes
}

\lstdefinelanguage{ACMPython}[]{Python}{
  morekeywords={with,as,match,case,Nonlocal,nonlocal}, % extend as needed
}
\lstdefinelanguage{ACMCpp}[]{C++}{
  morekeywords={alignas,alignof,consteval,constexpr,constinit,co_await,co_yield,co_return,concept,requires}
}
 
\definecolor{blue}{RGB}{37, 104, 185}    % Deep blue
\definecolor{green}{RGB}{43, 111, 109}   % Deep green
\definecolor{pink}{RGB}{186, 54, 100}    % Deep pink
\definecolor{purple}{RGB}{116, 70, 245}  % Deep purple
\definecolor{orange}{RGB}{166, 75, 23}   % Deep orange
\definecolor{grey}{RGB}{72, 72, 72}      % Deep grey
\definecolor{black}{RGB}{0, 0, 0}        % Black

\newcommand{\showcomments}{yes}
\newcommand\fixme[1]{
    \ifthenelse{\equal{\showcomments}{yes}}{\textcolor{red}{\small [FIXME: #1]}}{\ignorespaces}
}
\newcommand\ns[1]{
    \ifthenelse{\equal{\showcomments}{yes}}{\textcolor{purple}{\small [ns: #1~]}}{\ignorespaces}
}
\newcommand\wb[1]{
    \ifthenelse{\equal{\showcomments}{yes}}{\textcolor{purple}{\small [wb: #1~]}}{\ignorespaces}
}
\newcommand\zz[1]{
    \ifthenelse{\equal{\showcomments}{yes}}{\textcolor{blue}{\small [zz: #1]}}{\ignorespaces}
}
\newcommand\revise[1]{\textcolor{black}{#1}}

\AtBeginDocument{%
  }

\copyrightyear{2025}
\acmYear{2025}
\setcopyright{cc}
\setcctype{by}
\acmConference[MICRO '25]{58th IEEE/ACM International Symposium on Microarchitecture}{October 18--22, 2025}{Seoul, Republic of Korea}
\acmBooktitle{58th IEEE/ACM International Symposium on Microarchitecture (MICRO '25), October 18--22, 2025, Seoul, Republic of Korea}\acmDOI{10.1145/3725843.3756132}
\acmISBN{979-8-4007-1573-0/2025/10}

\begin{document}

% \input{cover-page}

% Reset page number starting from 1
\clearpage
\pagenumbering{arabic}

\title{Characterizing and Optimizing Realistic Workloads on a Commercial Compute-in-SRAM Device}
% \subtitle{\normalsize{MICRO 2025 Submission
    % \textbf{\#2096} -- Confidential Draft -- Do NOT Distribute!!}}
%\author{\normalsize{ISCA 2025 Submission
 %   \textbf{\#NaN} -- Confidential Draft -- Do NOT Distribute!!}}
\author{Niansong Zhang}
\affiliation{%
  \institution{Cornell University}
  \city{Ithaca}
  \state{NY}
  \country{USA}
}
\email{nz264@cornell.edu}

\author{Wenbo Zhu}
\authornote{This work was done during an internship at Cornell University.}
\affiliation{%
  \institution{University of Southern California}
  \city{Los Angeles}
  \state{CA}
  \country{USA}
}
\email{wenbozhu@usc.edu}

\author{Courtney Golden}
\affiliation{%
  \institution{MIT}
  \city{Cambridge}
  \state{MA}
  \country{USA}
}
\email{cgolden@csail.mit.edu}

\author{Dan Ilan}
\affiliation{%
  \institution{GSI Technology Inc.}
  \city{Tel Aviv}
  \country{Israel}
}
\email{dilan@gsitechnology.com}

\author{Hongzheng Chen}
\affiliation{%
  \institution{Cornell University}
  \city{Ithaca}
  \state{NY}
  \country{USA}
}
\email{hzchen@cs.cornell.edu}

\author{Christopher Batten}
\affiliation{%
  \institution{Cornell University}
  \city{Ithaca}
  \state{NY}
  \country{USA}
}
\email{cbatten@cornell.edu}

\author{Zhiru Zhang}
\affiliation{%
  \institution{Cornell University}
  \city{Ithaca}
  \state{NY}
  \country{USA}
}
\email{zhiruz@cornell.edu}


\input{sections/0-abstract}




\maketitle

\input{sections/1-intoduction}
\input{sections/2-background}
\input{sections/3-analytical}

\input{sections/4-optimizations}
\input{sections/5-evaluation}

\input{sections/6-related-work}

\input{sections/7-conclusion}

\input{sections/8-acknowledgements}

\bibliographystyle{ACM-Reference-Format}
\bibliography{main}

\end{document}

%% file: sections/0-abstract.tex
\begin{abstract}

Compute-in-SRAM architectures offer a promising approach to achieving higher performance and energy efficiency across a range of data-intensive applications. However, prior evaluations have largely relied on simulators or small prototypes, limiting the understanding of their real-world potential. In this work, we present a comprehensive performance and energy characterization of a commercial compute-in-SRAM device, the GSI APU, under realistic workloads. We compare the GSI APU against established architectures, including CPUs and GPUs, to quantify its energy efficiency and performance potential. 
We introduce an analytical framework for general-purpose compute-in-SRAM devices that reveals fundamental optimization principles by modeling performance trade-offs, thereby guiding program optimizations.

Exploiting the fine-grained parallelism of tightly integrated memory-compute architectures requires careful data management. We address this by proposing three optimizations: communication-aware reduction mapping, coalesced DMA, and broadcast-friendly data layouts. 
When applied to retrieval-augmented generation (RAG) over large corpora (10GB--200GB), these optimizations enable our compute-in-SRAM system to accelerate retrieval by 4.8$\times$--6.6$\times$ over an optimized CPU baseline, improving end-to-end RAG latency by 1.1$\times$--1.8$\times$. 
The shared off-chip memory bandwidth is modeled using a simulated HBM, while all other components are measured on the real compute-in-SRAM device.
Critically, this system matches the performance of an NVIDIA A6000 GPU for RAG while being significantly more energy-efficient (54.4$\times$-117.9$\times$ reduction). These findings validate the viability of compute-in-SRAM for complex, real-world applications and provide guidance for advancing the technology.

\end{abstract}

%% file: sections/1-intoduction.tex
\section{Introduction}

Compute-in-memory (CIM) holds the promise of being a highly energy-efficient approach to accelerating data-intensive applications by reducing memory access overhead through the integration of compute units within or near memory arrays. Among CIM approaches, compute-in-SRAM stands out for its compatibility with standard CMOS technology and potential to achieve high memory bandwidth.
The architectural and full-stack optimization of compute-in-SRAM systems continues to attract significant research interest. 
%Recent works have explored diverse designs for in-SRAM computing. Compute Caches~\cite{aga2017compute} repurposes cache elements as vector computational units using emerging bit-line SRAM technology, achieving a 1.9$\times$ performance gain and 2.4$\times$ energy savings for data-centric applications compared to a 32-byte SIMD CPU. EVE~\cite{al2023eve} introduces a bit-hybrid execution mechanism designed to enhance the efficiency of vector operations, achieving nearly 8$\times$ speedup compared to an out-of-order CPU. CAPE~\cite{caminal2021cape} presents a CMOS-based associative engine with significant programmability, providing an average speedup of 14$\times$, with peak speedups reaching up to 254$\times$ over an area-equivalent CPU.
%Additionally, specialized compute-in-SRAM accelerators, such as iMTransformer~\cite{laguna2022hardware}, TranCIM~\cite{tu2022trancim}, PICMA~\cite{zhang2022pimca}, and iMCAT~\cite{laguna2021memory}, target deep neural networks (DNNs) and transformer-based models, highlighting the potential for domain-specific acceleration.
%
Recent works propose diverse designs: Compute Caches~\cite{aga2017compute} repurpose cache elements as vector units using bit-line SRAM, delivering 1.9$\times$ speedup and 2.4$\times$ energy savings over a 32-byte SIMD CPU; EVE~\cite{al2023eve} uses a bit-hybrid execution mechanism to accelerate vector operations by nearly 8$\times$ versus an out-of-order CPU; and CAPE~\cite{caminal2021cape} offers a programmable CMOS associative engine, averaging 14$\times$ speedup with peaks up to 254$\times$ over an area-equivalent CPU. Specialized accelerators such as iMTransformer~\cite{laguna2022hardware}, TranCIM~\cite{tu2022trancim}, PICMA~\cite{zhang2022pimca}, and iMCAT~\cite{laguna2021memory} target deep neural networks (DNNs) and transformer models, highlighting the potential for domain-specific acceleration.

\input{tables/apu-overview}

\begin{figure}[t]
    \centering
    \includegraphics[width=\linewidth]{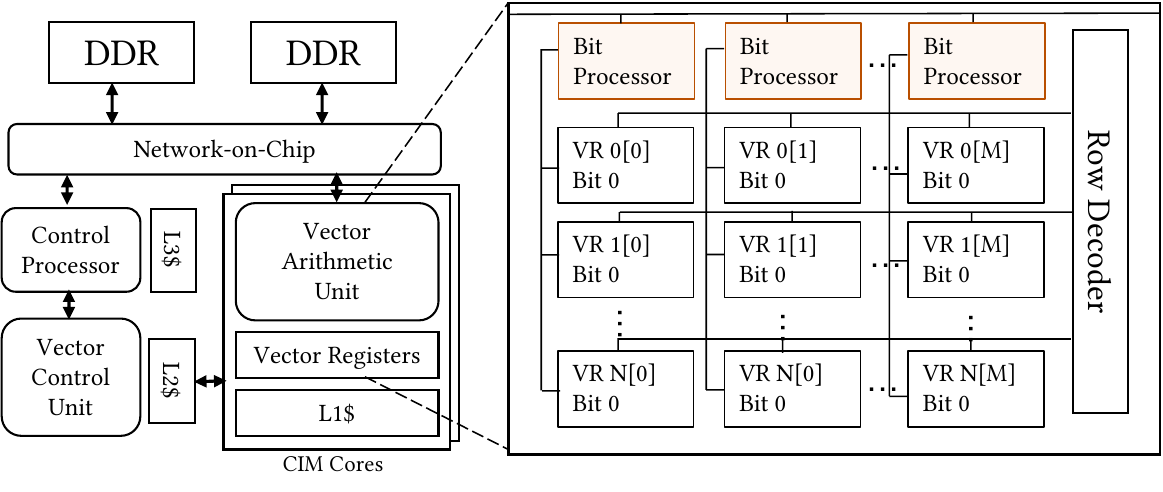}
    \caption{General-purpose compute-in-SRAM system model -- CAPE~\cite{caminal2021cape}, and GSI APU~\cite{gsiapu, gwennap2020gemini} follow the same system abstraction, with different vector arithmetic unit and SRAM cell implementations.}
    \vspace{-.2in}
    \label{fig:cim}
\end{figure}

Despite promising results, these architectures are primarily evaluated through instruction modeling and simulation~\cite{aga2017compute, zha2020hyper, laguna2022hardware, laguna2021memory, zhang2022pimca} or small-scale prototypes~\cite{tu2022trancim}, limiting insights into their practical, real-world effectiveness. This gap underscores the need for performance characterization of commercial compute-in-SRAM devices under realistic workloads, bridging the divide between theoretical promise and practical feasibility.

General-purpose compute-in-SRAM systems typically employ a SIMD vector processor abstraction~\cite{aga2017compute, caminal2021cape, zha2020hyper}. As illustrated in Fig.~\ref{fig:cim}, a common compute-in-SRAM architecture—adopted by systems like CAPE—abstracts the computation-enabled SRAM array as a vector processing engine. The GSI APU~\cite{gsiapu, gwennap2020gemini} aligns with this same abstraction, representing a commercial instance that provides a unique opportunity to evaluate the potential of compute-in-SRAM systems under realistic workloads and applications.

The GSI APU integrates 2 million bit processors at 500 MHz, delivering up to 25 TOPs for 8-bit addition~\cite{gsiapu}. Table~\ref{tab:apu-overview} compares its compute capacity, memory bandwidth, and power efficiency against CPUs, GPUs, and ASIC accelerators, showing strong potential for data-intensive workloads. Fully exploiting this is difficult: the APU uses a 32,768-element vector processor with column-wise integrated compute and storage, offering TB/s on-chip bandwidth but limiting memory access within a vector register (VR). For instance, reductions across a VR are unsupported, and intra-VR group operations are about 10$\times$ slower than inter-VR operations.

\begin{figure}[t]
    \centering
    \includegraphics[width=.9\linewidth]{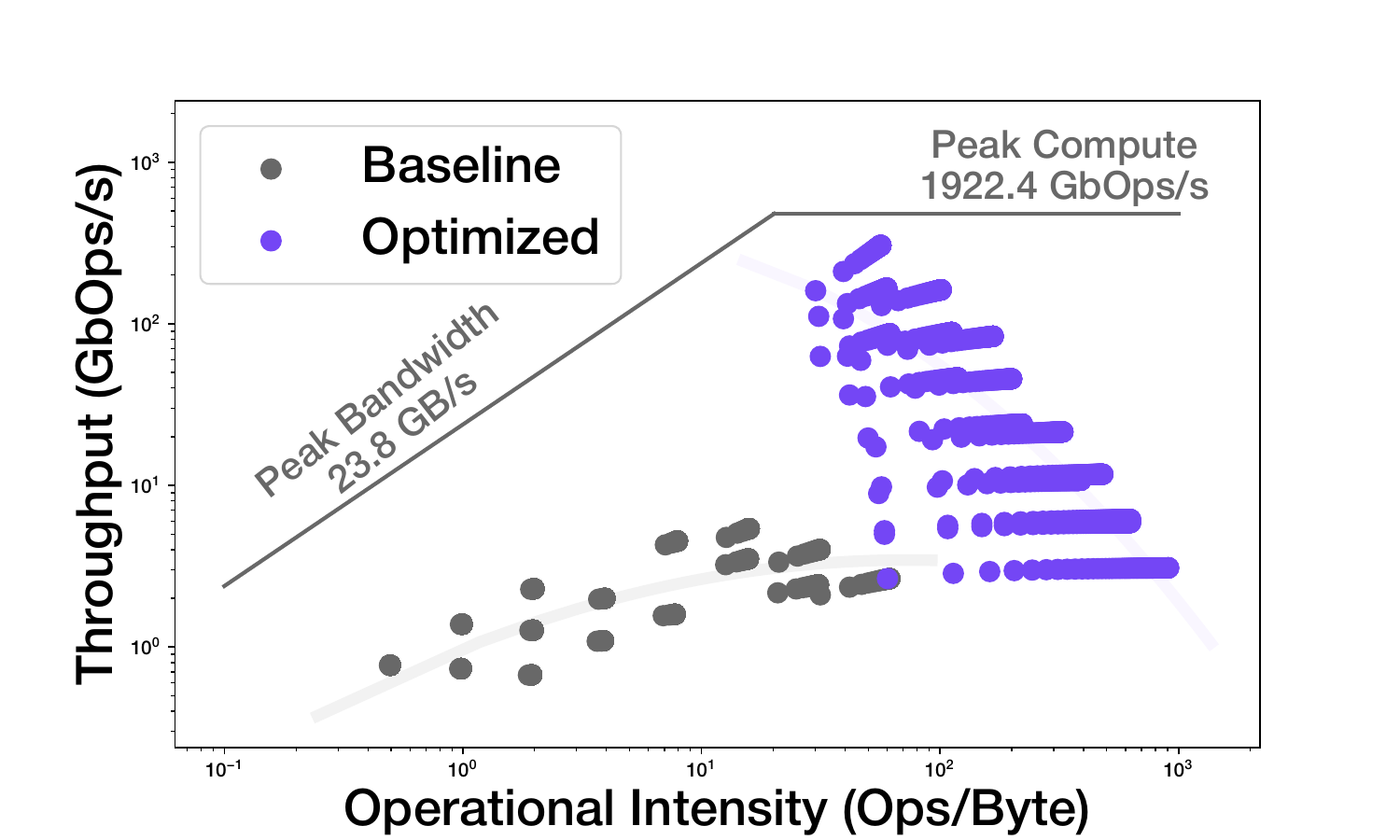}
    \caption{Performance of various matrix multiplication kernels on GSI APU, w/o data movement and data layout optimizations.}
    \label{fig:roofline}
    \vspace{-.1in}
\end{figure}

Figure~\ref{fig:roofline} shows a roofline model of different matrix multiplication kernels on the GSI APU\footnote{The peak computational bound is profiled for 16-bit unsigned multiplication and accumulation operations.}. The baseline approach, implementing a vectorized inner-product algorithm, does not account for data movement or layout overheads, resulting in suboptimal performance. However, with tailored data optimizations, performance approaches the compute roof with higher operational intensity. 

This observation highlights a broader insight about compute-in-SRAM devices: despite performing computation directly within memory, these systems can still be bottlenecked by memory bandwidth if data movement is not carefully managed. To further analyze this issue, we develop an analytical framework that exposes the underlying performance limits.
In this work, we propose three key optimizations to realize the potential of the compute-in-SRAM systems:  communication-aware reduction mapping, coalesced DMA operations, and broadcast-friendly data layouts.

Furthermore, we evaluate compute-in-SRAM systems for Retrieval-Augmented Generation (RAG) in large language model (LLM) inference, demonstrating its suitability for this workload. We also compare its performance and energy efficiency against CPU and GPU to highlight its advantages.
Our contributions are as follows:

\begin{itemize}[left=0pt]
\item We present the first comprehensive evaluation of a commercial compute-in-SRAM device using realistic workloads. Specifically, we assess the GSI APU—a commercial instance of a general-purpose compute-in-SRAM device—using the Phoenix benchmark, matrix multiplication, and retrieval-augmented generation (RAG) workloads. We compare its performance and energy efficiency against established architectures, including an Intel Xeon Gold CPU and an NVIDIA A6000 GPU.

% \item We develop an analytical framework to identify optimization opportunities, providing insights into performance and efficiency trade-offs. This framework informs the design of next-generation in-SRAM computing architectures and guides future compute-in-SRAM optimizations.

\item \revise{We develop a flexible analytical framework that identifies optimization opportunities and supports architectural design space exploration by enabling the tuning of key design parameters. This framework informs the design of next-generation in-SRAM computing architectures.}

\item We propose three key optimizations targeting data movement and layout to exploit the unique characteristics of ultra-long vector compute-in-SRAM architectures. Applied to the RAG workload, these optimizations reduce retrieval latency by up to 6.6$\times$ compared to an optimized CPU baseline, yielding up to 1.8$\times$ end-to-end speedup and matching the latency of GPU-based systems while consuming 1\% of the energy. On Phoenix, the optimized APU achieves a 41.8$\times$ speedup over CPU.
\end{itemize}

%% file: tables/apu-overview.tex
\begin{table}[t]
\centering
\caption{Comparison of GSI APU~\cite{gsiapu, gwennap2020gemini}, Intel Xeon 8280, NVIDIA A100, and Graphcore IPU.}
\resizebox{\linewidth}{!}{
\begin{tabular}{>{\bfseries}lcccc}
\toprule
 & \textbf{GSI APU} & \textbf{Xeon 8280} & \textbf{NVIDIA A100} & \textbf{Graphcore} \\
\hline
Compute Cores & 2 million$\times$1 bit & 28$\times$2$\times$512 bits & 104$\times$4,096 bits & 1,216$\times$64 bits \\
Tech node & 28 nm & 14 nm & 7 nm & 7 nm \\
Compute Speed & 500 MHz & 2.7 GHz & 1.4 GHz & 1.6 GHz \\
Peak Compute & 25 TOPS & 10 TOPS & 75 TOPS & 16 TOPS \\
On-Chip Memory & 12MB L1 & 38.5MB L3 & 40MB L2 & 300MB L1 \\
Mem. Bandwidth & 26 TB/s & 1 TB/s & 7 TB/s & 16 TB/s \\
Power & 60W TDP & 205W TDP & 400W TDP & 150W TDP \\
\bottomrule
\end{tabular}
\label{tab:apu-overview}
}
\end{table}

%% file: sections/2-background.tex
\section{GSI APU Architecture}

\subsection{Architecture and Microarchitecture}

\begin{figure*}[!htbp]
    \centering
    \begin{tabular}{cc}
        \includegraphics[width=0.3\linewidth]{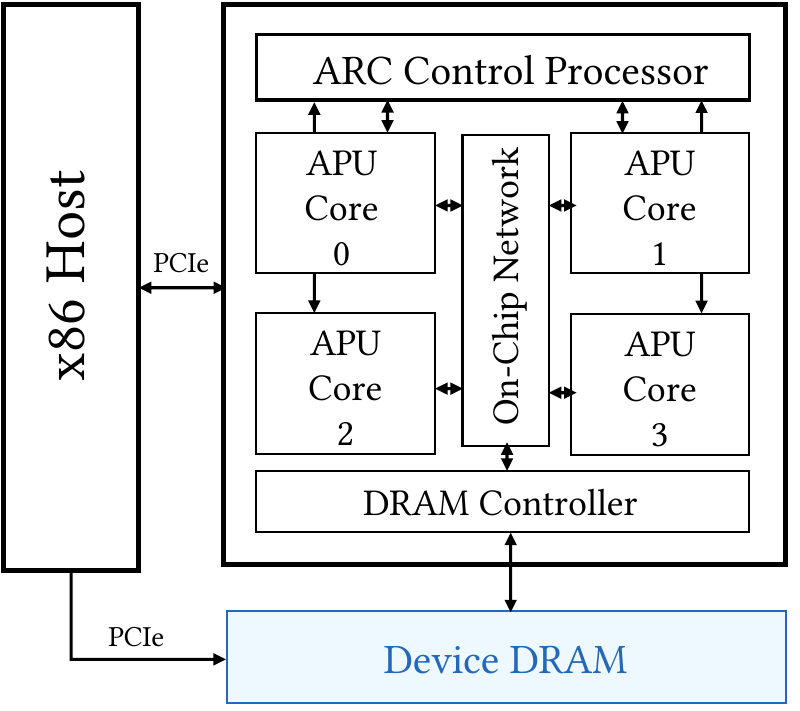} & 
        \includegraphics[width=0.66\linewidth]{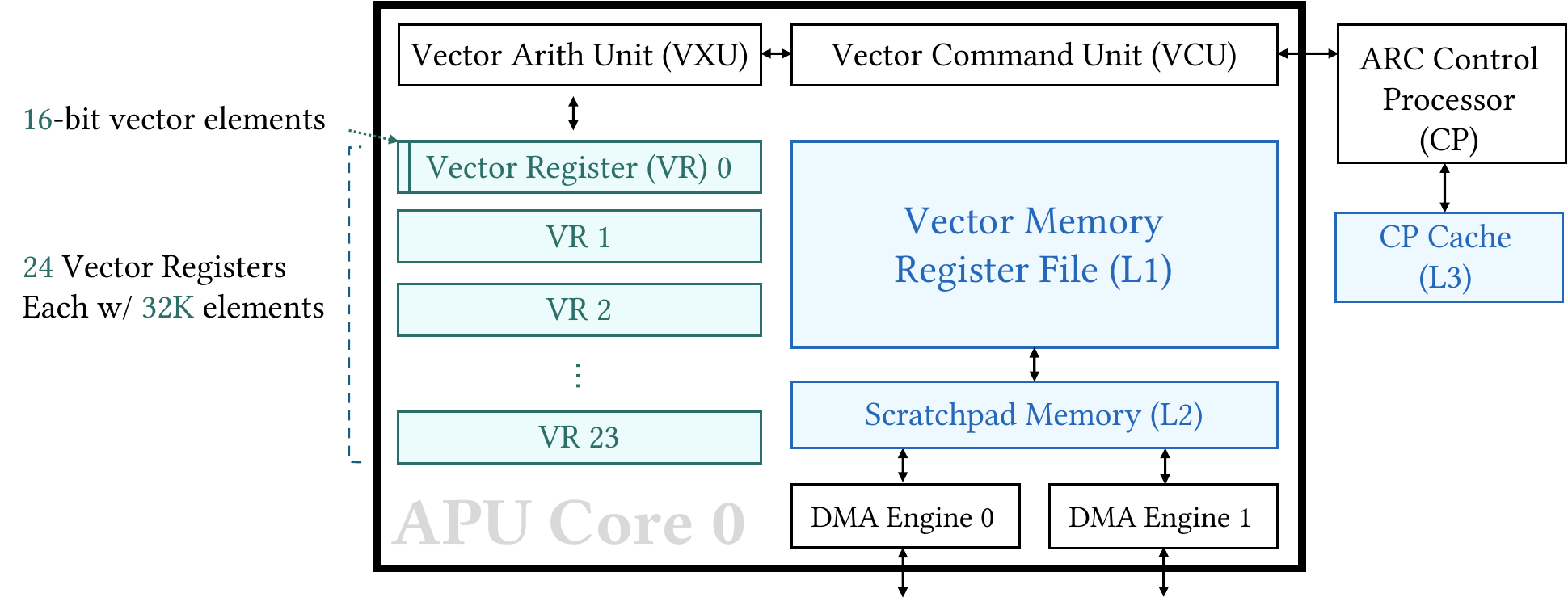} \\
        (a) System overview. & (b) APU core logic view. \\
    \end{tabular}
    \caption{The GSI APU system and APU core logic view, the memory hierarchy is highlighted in \textcolor{blue}{blue}. The APU consists of a control processor, four APU cores, and a four-level memory hierarchy. Each core has 24 \textcolor{green}{vector registers} (VR), and each VR has 32768 elements.}
    \label{fig:arch}
\end{figure*}

% \begin{figure*}[!htbp]
%     \centering
%     \begin{tabular}{cc}
%         \includegraphics[width=0.65\linewidth]{figs/physical_banks.pdf} & 
%         \includegraphics[width=0.3\linewidth]{figs/bit_processor.pdf} \\
%         (a) APU core physical bank organization. & (b) Bit processor architecture. \\
%     \end{tabular}
%     \caption{The physical bank organization of one GSI APU core and the \textcolor{orange}{bit processor} architecture. The data is stored in a bit-slice fashion, each column of each bit-slice contains a single-bit read latch and associated read and write logic.}
%     \label{fig:uarch}
% \end{figure*}

\begin{figure*}[!htbp]
    \centering
    \includegraphics[width=\linewidth]{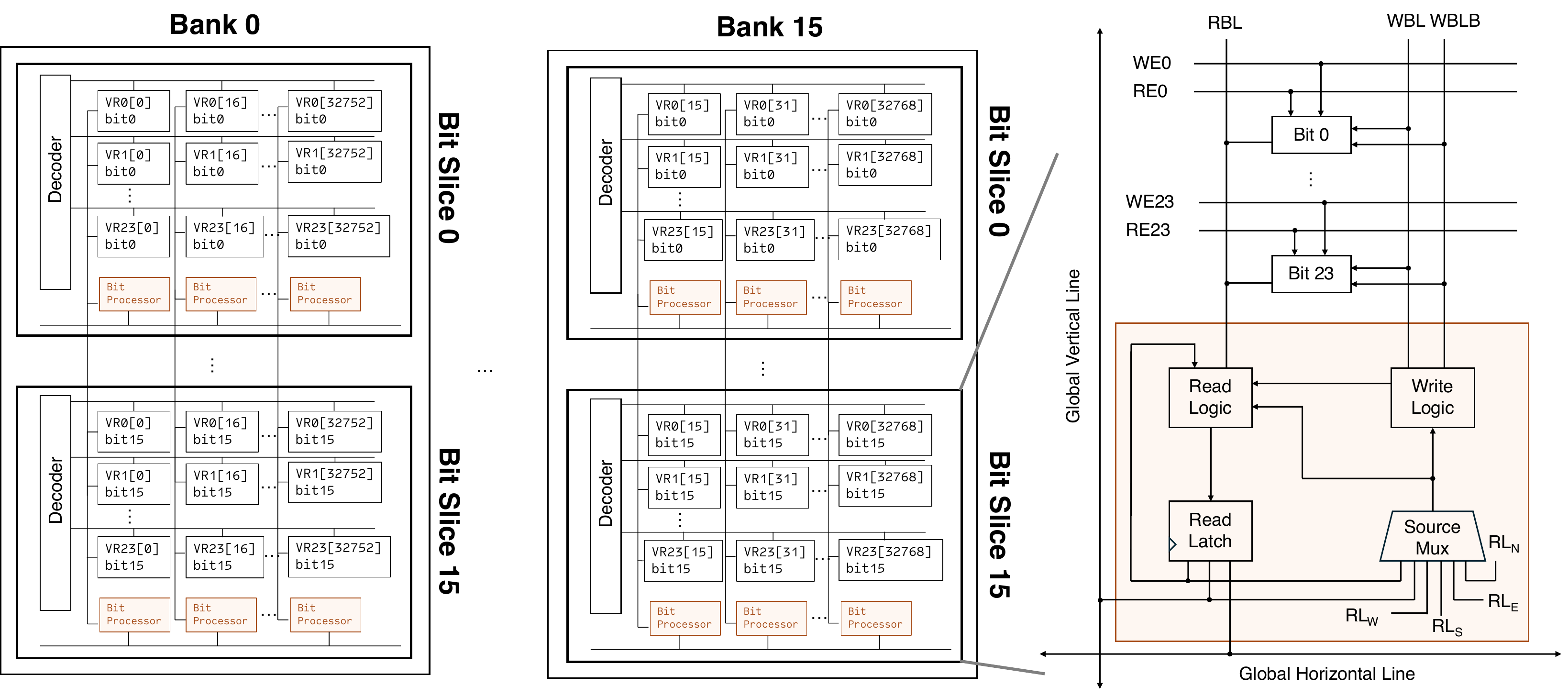} \\
    \begin{tabular}{p{0.65\linewidth} p{0.3\linewidth}}
        \parbox{\linewidth}{\centering (a) APU core physical bank organization.} &
        \parbox{\linewidth}{\centering (b) Bit processor architecture.}
    \end{tabular}
    \caption{The physical bank organization of one GSI APU core and the \textcolor{orange}{bit processor} architecture. The data is stored in a bit-slice fashion, each column of each bit-slice contains a single-bit read latch and associated read and write logic.}
    \label{fig:uarch}
\end{figure*}

In this section, we provide an overview of the APU's architecture and microarchitecture. As shown in Fig.~\ref{fig:arch}(a), the APU platform comprises a standard x86-64 host CPU and a four-core APU chip, connected via PCIe and sharing a DDR4 DRAM. Each APU core functions as a vector engine, processing 32K-element vectors of 16-bit data, as shown in Fig.~\ref{fig:arch}(b). The 32-bit ARC control processor (CP) issues vector commands to the Vector Command Unit (VCU), and the VCU decodes the vector command to microcode operations to load/store vectors to the Vector Registers (VRs) and perform arithmetic computations. 

The memory hierarchy is highlighted in \textcolor{blue}{blue} in Fig.~\ref{fig:arch}, which consists of a 16 GB device DRAM, a 1 MB control processor (CP) cache (L3), a 64KB scratch pad memory (L2), and a 3 MB vector memory register file (L1). 
The DDR memory is shared by four APU cores, and each core has its private L2 and L1 memory. The L2 scrachpad memory serves as a DMA buffer to contain one 32K-element, 16-bit vector. The L1 memory is organized as 48 "background" registers as additional storage to the computation-enabled VRs.

Figure~\ref{fig:uarch}(a) shows the physical bank organization of the VRs. The 24 VRs are striped across 16 physical banks, and each bank contains 2048 16-bit elements.
Within one physical bank, the data is stored in a bit-slice fashion, where each bit-slice contains one bit for all 24 VRs. Each column of each bit-slice integrates a bit processor with 24 custom 12 T SRAM cells. 
The bit processor microarchitecture is shown in Fig.~\ref{fig:uarch}(b). The bit processors are collectively equivalent to the vector arithmetic unit (VXU). The read logic can perform AND, OR, and XOR on two or more operands, including data from \revise{the} read bit-line (RBL), the read latch (RL), the global vertical line, \revise{the} global horizontal line, and the RLs of bit processors to its north ($RL_N$), south ($RL_S$), east ($RL_E$), or west ($RL_W$). 
The global horizontal line connects all bit processors in the same row, while the global vertical line connects those in the same column. Each line includes a 1-bit latch: the global horizontal latch (GHL) and the global vertical latch (GVL).
If multiple values are read to GHL simultaneously, an OR operation is performed before storing the result to the latch. For GVL, it performs an AND on the multiple values. The write logic updates the SRAM cells through the write bit-line (WBL) or its negation (WBLB). By default, bit processors in all bit-slices are issued the same micro operation. However, a 16-mask can be used to perform the operations on a subset of the bit slices. The VRs, RL, GHL, and GVL are the microarchitectural states. The operations on the microarchitectural states are listed in \revise{Table~\ref{tab:state}}.

\input{tables/apu-arch}

\subsubsection{Arithmetic Operations and Data Types}

Unlike bit-serial architectures that process only one bit at a time, the APU supports both bit-serial arithmetic and bit-parallel boolean operations. This flexibility is achieved through the bit-slice bank organization shown in Fig.~\ref{fig:uarch}(a), allowing all bits of a VR to be accessed simultaneously by the bit processors. The APU natively supports 16-bit signed and unsigned integers, 16-bit IEEE floating point, and a custom GSI floating point format with a 6-bit exponent and a 9-bit mantissa.

\subsubsection{Data Movement}
The APU supports both direct memory access (DMA) and programmable I/O (PIO). As shown in Fig.~\ref{fig:arch}(b), each APU core is equipped with two parallel DMA engines that transfer contiguous data in 512-byte chunks, enabling high memory bandwidth for VR transfers within the memory hierarchy. For random access or single-element extraction from the VR, the ARC control processor can perform these operations using PIO. For DRAM~$\leftrightarrow$~L3 transfers, both DMA and PIO can be used, whereas for DRAM~$\leftrightarrow$~L2, only DMA is available.

\noindent \textbf{DRAM $\leftrightarrow$ L3, DRAM $\leftrightarrow$ L2}:
For these types of data movement, data layout transformations can be applied.  With DMA, the source and target 512-byte chunk addresses can be programmed to enable contiguous, strided, and duplicated data layout transformations. PIO enables arbitrary data layout transformations, though with lower bandwidth compared to DMA.

\noindent \textbf{L2 $\leftrightarrow$ L1, L1 $\leftrightarrow$ VR}: 
For these types of data movement, data layout transformations are not supported. Data is transferred at the granularity of an entire vector, meaning only full VR loads/stores (32K by 16-bit) are possible.

\noindent \textbf{L3 $\leftrightarrow$ VR}: 
PIO enables direct data transfers between L3 and VRs via a response FIFO (RSP FIFO). It supports serial retrieval (get) from VR and parallel insertion (set) into VR. The CP can broadcast scalars or immediate values to entire VRs or masked elements, while retrieval from VR is limited to one element at a time.

\noindent \textbf{Inter-VR vs. intra-VR}\revise{:}
Due to the bit-slice organization, element-wise data movement between VRs can be performed efficiently, as all elements and bits can be processed in parallel by the bit processors. However, intra-VR data movements, such as vector shifts or bank copies, depend on the GHL or RSP FIFO and thus cannot be fully parallelized.

\subsubsection{Implications on Data Layout}
The differing costs of inter- and intra-VR data movement impact how data layout in the memory hierarchy affects performance in several ways: (1) For device DRAM and L3, data layout influences the bandwidth of data movement. When data is contiguous or has a regular stride, DMA offers higher bandwidth than PIO. (2) Data layout within the VR also affects data movement time. If computation results are contiguous within the VR, DMA can efficiently transfer them back to L1, L2, and then device DRAM. However, if they are scattered, PIO must be used to move them sequentially. (3) Data layout in the VR impacts computational efficiency. For instance, a reduction operation can be mapped to either inter-VR or intra-VR operations, depending on the data layout in the VR. As discussed in Section~\ref{sec:data_movement}, intra-VR data movement is more costly than inter-VR movement, making intra-VR reductions more expensive due to data movement overhead.

\input{code/programming_model}

\subsection{Programming Model}

The APU uses a host-accelerator programming model, where an x86 host manages kernel execution, shared memory, and kernel invocation on the APU device. Fig.~\ref{fig:main_code} shows this model using a simple vector addition example to demonstrate host-device interaction.

\subsubsection{Host Program}
The host program, written in C, manages kernel invocation, shared DRAM (L4) memory allocation and deallocation, and data transfers between the host and device memory. Fig.~\ref{fig:main_code}(a) shows a snippet of the host-side code. Initialization of the calling context and input data is omitted for simplicity. The host and device communicate through a program command structure, detailed in lines L1--L9. Memory management, including device memory allocation, data movement, and kernel invocation, is handled by the GSI GDL library, a memory management library from GSI.

\subsubsection{Device Program}
The device program, also in C, runs on the APU control processor and uses general-purpose control flow statements. The system macro \texttt{GAL\_TASK\_ENTRY\_POINT} defines the entry point of the device program, extracts the data structure from the command, and calls the \texttt{vec\_add} function. The device program manages data transfers from device memory to L1 memory and performs vector computations using Vector Registers (VRs).
Vector processing uses the GSI Vector Math Library (GVML), which provides functions for vector operations, including arithmetic, logical, bitwise, trigonometric, and min/max operations. Once computations are complete, the device program transfers data back to device memory.

The GVML library is implemented using APU microcode instructions. APU microcode instructions directly \revise{operate} on the microarchitectural states listed in Table.~\ref{tab:state}. An APU programmer can implement a different vector abstraction with microcode instructions. For example, Golden et al.~\cite{golden2023supporting} implemented a RISC-V vector abstraction using APU microcode. In this work, we use the abstractions provided by GVML to focus on optimizing performance through these data movement and computation operations.

%% file: tables/apu-arch.tex
\begin{table}[t]
\centering
\caption{Microarchitectural state and operations on state.}
\resizebox{0.9\linewidth}{!}{
\begin{tabular}{l l}
\toprule
\textbf{State} & \textbf{Description} \\ \midrule
\textbf{RL} & read latch \\
\textbf{GVL} & global vertical latch \\
\textbf{GHL} & global horizontal latch \\
\textbf{VR[$i$]} & vector register source $i$ \\
\bottomrule
\toprule
\textbf{Operations} & \textbf{Description} \\ \midrule
\textbf{RL = VR[$vrs0$]} & read VR \\
\textbf{RL = VR[$vrs0, vrs1$]} & read and bitwise AND of two VRs \\
\textbf{RL = $L$} & read value from a source latch \\
\textbf{RL = VR[$vrs0$] $\,op\,$ $L$} & operate on a VR and a latch \\
\textbf{RL $\,op=\,$ VR[$vrs0$]} & operate on current RL and a VR \\
\textbf{RL $\,op=\,$ $L$} & operate on current RL and a latch \\
\textbf{RL $\,op=\,$ VR[$vrs0$] $\,op\,$ $L$} & operate on RL, a VR, and a latch \\
\textbf{VR[$vrs0$] = $L$} & write to VR from source latch \\
\bottomrule
\end{tabular}
}
\label{tab:state}
\end{table}

%% file: code/programming_model.tex
\begin{figure}[!htbp] % Use figure* for spanning two columns if needed
    \centering

    % First code block
    \begin{subfigure}[t]{0.45\textwidth} % Adjust width to fit one column
        \centering
        \includegraphics[width=\linewidth]{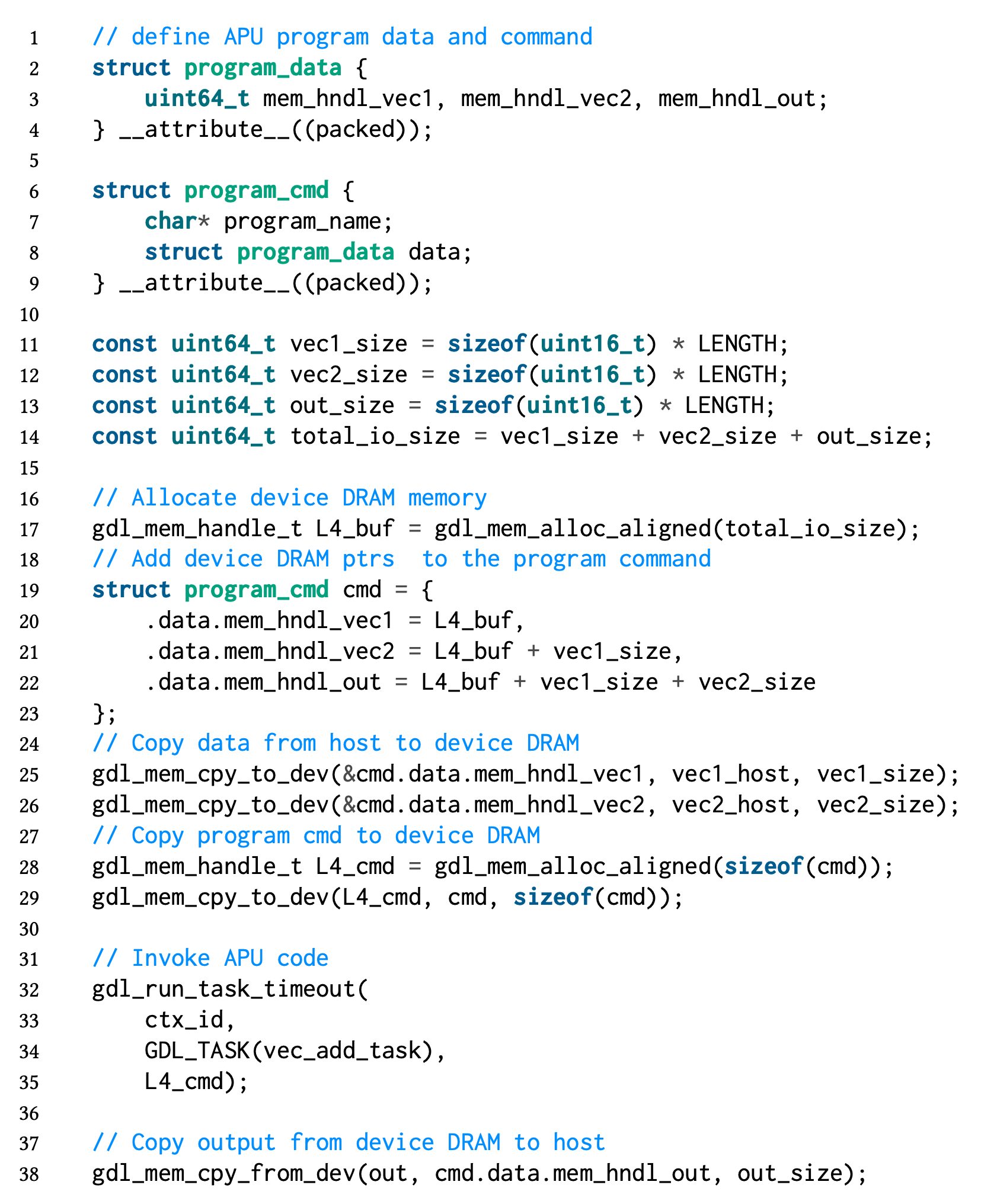}
        \caption{APU host code.}
        \label{fig:code1}
    \end{subfigure}
    \hfill % Add horizontal spacing between the subfigures
    \vspace{10pt}

    % Second code block
    \begin{subfigure}[t]{0.45\textwidth} % Adjust width to fit one column
        \centering
        \includegraphics[width=\linewidth]{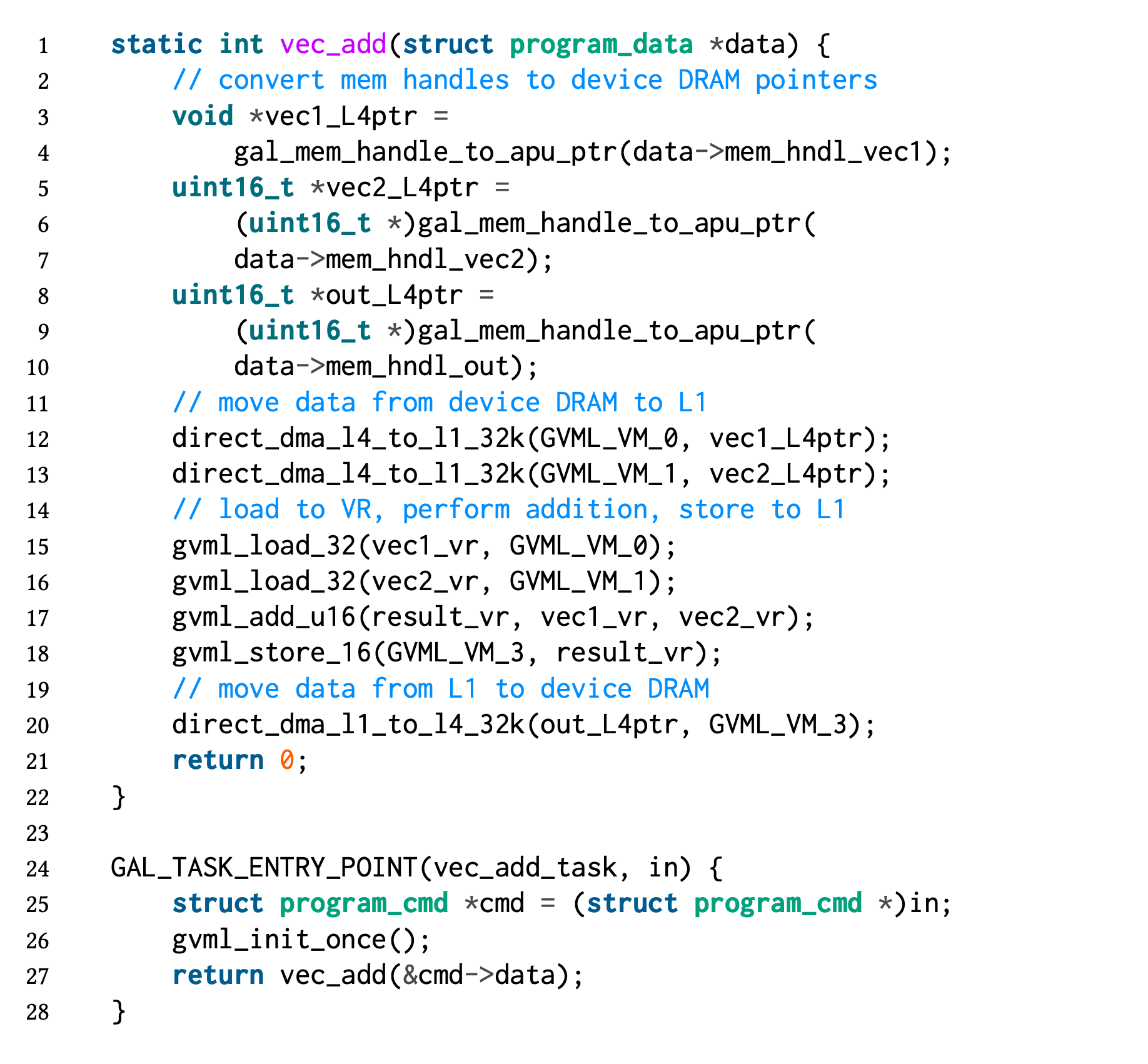}
        \caption{APU device code.}
        \label{fig:code2}
    \end{subfigure}

    % Main caption for the entire figure
    \caption{Simple vector addition code demonstrating APU programming model.}
    \label{fig:main_code}
    \vspace{-.2in}
\end{figure}

%% file: sections/3-analytical.tex
\section{Analytical Framework}

We propose a \revise{flexible} analytical framework to model performance characteristics of compute-in-SRAM platforms. This framework parameterizes critical architectural factors, including computation latency, data movement bandwidth, and communication patterns with potentially non-uniform costs. Such generalization enables applicability across various compute-in-SRAM architectures, aiding in performance analysis and optimization strategies beyond specific device implementations.

\subsection{Applicability and Assumptions}
This analytical framework targets compute-in-SRAM system models as illustrated in Fig.~\ref{fig:cim}. The model assumes a PCIe-based accelerator with a multi-level memory hierarchy and a vector processor abstraction, where data movement costs are non-uniform across memory levels and within vector registers. 
While the framework is validated using the GSI APU, it is not limited to this device. It can be extended to other compute-in-SRAM platforms that follow the same system model by deriving the necessary parameters through profiling.

Table~\ref{tab:notations} summarizes notations used throughout the framework. Tables~\ref{tab:analytical-memory} and~\ref{tab:analytical-compute} provide generic models of latency for data movement and computation operations, respectively. \revise{Framework validation against measured latencies on a real device is discussed in Section~\ref{sec:validate}.}

\input{tables/notations}

\subsection{Data Movement}
\label{sec:data_movement}

\input{tables/analytical-memory}

Effective data movement is crucial for compute-in-SRAM systems, particularly in data-intensive applications. Below, we discuss key data movement paradigms typically supported by these architectures.

\subsubsection{DMA Transfers}

Direct Memory Access (DMA) operations facilitate efficient bulk data transfers within compute-in-SRAM platforms. DMA latency generally scales linearly with transfer size, captured by the model $T_{\text{DMA}} = d/BW + T_{\text{init}}$, where $d$ is data size, $BW$ is bandwidth, and $T_{\text{init}}$ is a fixed initialization overhead. While DMA provides high throughput for continuous data movement, off-chip memory bandwidth constraints can limit performance for very large data sizes.

\subsubsection{Programmable I/O (PIO)}

PIO enables fine-grained, irregular data access patterns. The latency of PIO transfers typically scales with the number of individual load or store operations, modeled as $T_{\text{PIO}} = n \cdot T_{\text{access}}$, where $n$ is the operation count. Though flexible, PIO incurs higher overhead compared to DMA, making it suited primarily for non-contiguous or sparse data transfers.

\subsubsection{Indexed Lookup and Element-wise Operations}

Indexed lookups handle irregular, scatter-like data transfers from higher memory levels to local vector registers (VR). The lookup latency grows proportionally with table size ($\sigma$), formulated as $T_{\text{lookup}} = C \cdot \sigma + T_{\text{init}}$, highlighting the necessity for careful indexing and data layout optimization.
Element-wise copy operations, such as scalar broadcasting and VR-to-VR transfers, execute efficiently due to parallel hardware mechanisms, typically exhibiting constant-time latencies. Such operations are essential for data initialization and broadcasting in parallel workloads.

\subsubsection{Vector Register (VR) Shifts}

Intra-vector register shifts rearrange data locally within VRs without accessing external memory, incurring latency proportional to the shift magnitude, modeled by $T_{\text{shift\_e}} = C \cdot k$. Minimizing intra-VR shifts through optimized data layouts can significantly improve overall performance.

\subsection{Computation}

\input{tables/analytical-compute}

\input{code/analytical_framework}

On compute-in-SRAM platforms, vectorized arithmetic, logical, and comparison operations typically execute in constant time due to their inherent parallel execution. Therefore, we summarize their notation and provide representative latency measurements obtained from the GSI APU in Table~\ref{tab:analytical-compute}.

Reduction operations aggregate elements within vector registers, such as summation or finding extrema. Such operations often employ subgroup-based hierarchical reduction strategies to exploit parallelism. However, due to hardware constraints, inter-subgroup reductions may have non-linear costs and can be significantly higher than intra-subgroup operations. A generic cost model for subgroup-based reductions can be expressed as:

\begin{equation}
\begin{aligned}
    T_{\text{sg\_add}}(r, s) &= p_3 (\log_2{s})^3 + p_2 (\log_2{s})^2 + p_1 \log_2{s} + p_0, \\
    p_3 &= \alpha_3 \cdot \log_2{r} + \beta_3, \quad
    p_2 = \alpha_2 \cdot \log_2{r} + \beta_2, \\
    p_1 &= \alpha_1 \cdot \log_2{r} + \beta_1, \quad
    p_0 = \alpha_0 \cdot \log_2{r} + \beta_0.
\end{aligned}
\label{eq:sg_add_equations}
\end{equation}

\noindent The cubic term emerges due to the multi-level shifting, alignment, and accumulation operations inherent in hierarchical reductions, whose complexity grows non-linearly as subgroup size increases. Using logarithms (\(\log_2{s}\) and \(\log_2{r}\)) in the model is natural since these operations typically organize data aggregation in stages that halve the subgroup size at each step, indicating a logarithmic relationship. The polynomial coefficients \(p_0, p_1, p_2, p_3\) themselves depend logarithmically on the group size (\(r\)), parameterized by experimentally determined constants \(\alpha_i, \beta_i\). This generalized formulation allows modeling of complex, non-linear hardware behavior common in hierarchical reduction operations.

\begin{figure*}[htbp]
    \centering
    \begin{tabular}{@{}m{0.25\linewidth}<{\centering}@{}m{0.35\linewidth}<{\centering}@{}m{0.35\linewidth}<{\centering}@{}}
        \includegraphics[width=\linewidth]{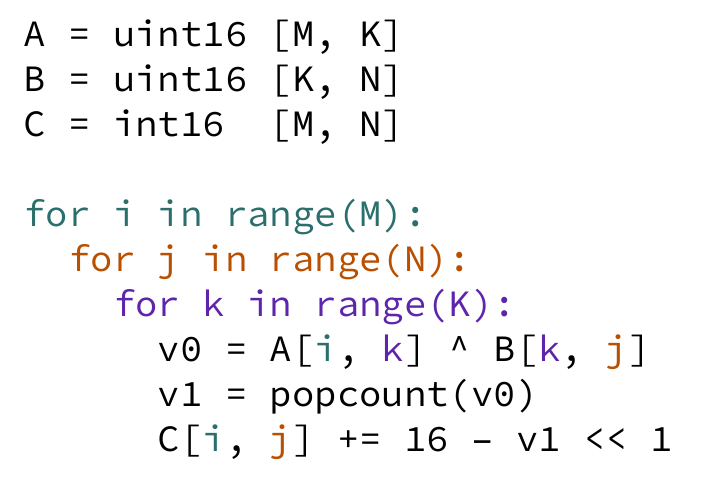} & 
        \includegraphics[width=\linewidth]{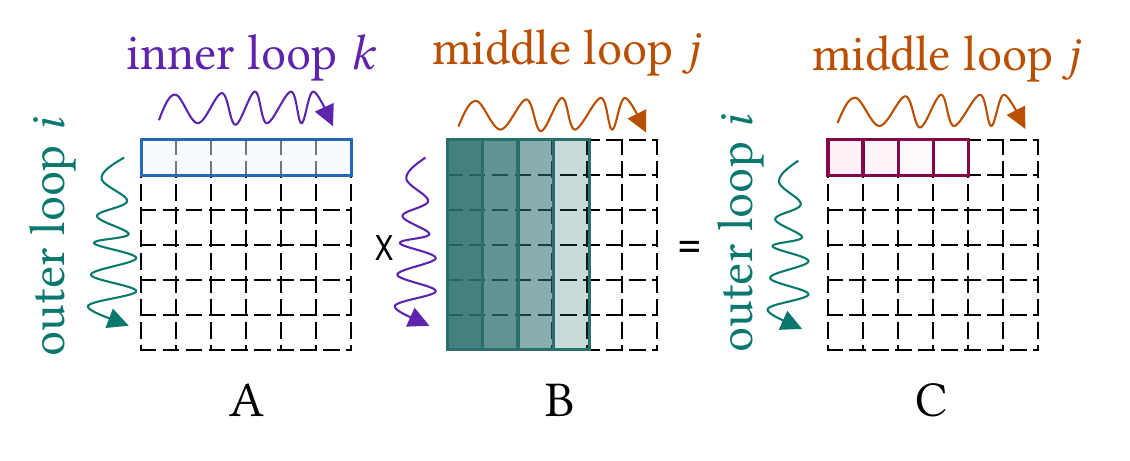} & 
        \includegraphics[width=\linewidth]{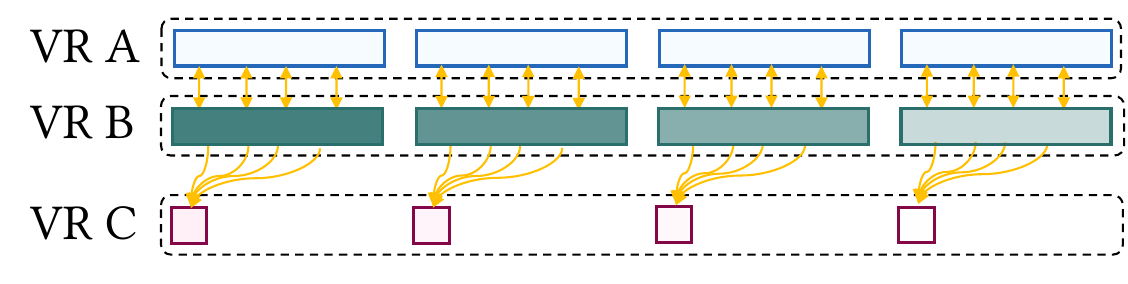} \\
        (a) Binary matrix multiply. & (b) Inner-product data access. & (c) Data layout on vector registers. \\
    \end{tabular}
    \vspace{-.1in}
    \caption{Motivating example: binary matrix multiply implemented as an inner-product algorithm on APU.}
    \label{fig:matmul}
\end{figure*}

\subsection{Framework Implementation}
\label{sec:framework_impl}
\revise{
We have developed a Python function library that mirrors the interface of the GSI-provided C++ API, enabling programmers to model arbitrary APU programs. The analytical framework interprets these programs and reports the total latency. Fig.~\ref{fig:analytical_impl} shows a code snippet modeling the \texttt{Histogram} application from the Phoenix benchmark suite~\cite{ranger2007evaluating}.
}

\subsection{Framework Implications}

Our analytical framework highlights general performance trends across compute-in-SRAM architectures. Specifically, element-wise computations exhibit low latency and efficiently exploit parallel hardware. Conversely, large-scale reduction operations and certain intra-vector data movements can become significant bottlenecks. DMA transfers outperform PIO for bulk movements but lack flexibility for sparse data access. Thus, achieving high efficiency on compute-in-SRAM platforms requires careful optimization of data movement strategies, data layouts, and computational structures aligned with underlying hardware characteristics.

%% file: tables/notations.tex
\begin{table}[t]
\centering
\caption{Notations}
\vspace{-.1in}
\label{tab:notations}
\resizebox{0.9\linewidth}{!}{
\begin{tabular}{crcr}
\toprule
\textbf{Notation}  & \textbf{Description}      & \textbf{Notation}& \textbf{Description} \\ \midrule
   $d$             & Data size in bytes        & BW           & Memory bandwidth   \\ 
   $r$             & VR group size             & $s$            & Subgroup size  \\ 
   $\sigma$        & Lookup table size         & $C$            & Constant \\ \bottomrule
\end{tabular}
}
\end{table}

%% file: tables/analytical-memory.tex
\begin{table}[t]
  {
  \caption{Data movement analytical framework}
  \vspace{-.1in}
  \label{tab:analytical-memory}
  \resizebox{\linewidth}{!}{
  \begin{tabular}{lllr}
  \toprule
    \multirow{2}{*}{\textbf{Operation}} & \multirow{2}{*}{\normalfont\textbf{Description}} & \multicolumn{2}{c}{\textbf{Execution Time (cycles)}} \\
                                        &                                                 & \multicolumn{2}{l}{\textbf{Analytical}\hspace{0.45in}\textbf{Meas.}} \\\midrule
    \tt{dma\_l4\_l3}            & L4$\rightarrow$L3 DMA                        & $d/\text{BW} + T_{\text{init}}$ &   $0.19d+41164$ \\ 
    \tt{dma\_l4\_l2}            & L4$\rightarrow$L2 DMA                        & $d/\text{BW} + T_{\text{init}}$ &   $0.63d+548$ \\
    \tt{dma\_l2\_l1}            & L2$\rightarrow$L1 DMA, 16-bit $\times$ 32K   & $T_{\text{l2}\rightarrow \text{l1}}$ & 386 \\
    \tt{dma\_l4\_l1}            & L4$\rightarrow$L1 DMA, 16-bit $\times$ 32K   & $T_{\text{l4}\rightarrow \text{l1}}$ & 22272 \\
    \tt{dma\_l1\_l4}            & L1$\rightarrow$L4 DMA, 16-bit $\times$ 32K   & $T_{\text{l1}\rightarrow \text{l4}}$ & 22186 \\
    \tt{pio\_ld}                & PIO load, L4$\rightarrow$VR                  & $n \cdot T_{\text{pio\_ld}}$          & $57n$   \\
    \tt{pio\_st}                & PIO store, VR$\rightarrow$L4                 & $n \cdot T_{\text{pio\_st}}$          & $61n$   \\
    \tt{lookup}                 & Lookup L3 w/ index VR                        & $C \cdot \sigma + T_{\text{init}}$   & $7.15\sigma + 629$  \\
    \hline
    \tt{load, store}            & VR$\leftrightarrow$L1 load store             & $T_{\text{ld}}$, $T_{\text{st}}$           &   29 \\ 
    \tt{cpy}                    & VR$\leftrightarrow$VR element-wise copy       & $T_{\text{cpy}}$                    &   29 \\ 
    \tt{cpy\_subgrp}            & Copy VR subgroup to group                    & $T_{\text{cpy\_sgp}}$               &   82 \\
    \tt{cpy\_imm}               & Broadcast an immediate to VR                 & $T_{\text{cpy\_imm}}$               &   13 \\   
    \tt{shift\_e(k)}            & Shift VR entries to head/tail by $k$         & $C\cdot k$                          &   $373k$ \\
    \tt{shift\_e(4k)}           & Intra-bank shift VR entries by $4\cdot k$    & $C + k$                      &   $8 + k$ \\
  \bottomrule
  \end{tabular}\par}
  }
  \raggedright\footnotesize
* In the analytical framework, we refer to the device DRAM as L4 memory.
\end{table}

%% file: tables/analytical-compute.tex
\begin{table}[t]
  {
  \caption{Computation analytical framework}
  \vspace{-.1in}
  \label{tab:analytical-compute}
  \resizebox{\linewidth}{!}{
  \begin{tabular}{lllr}
  \toprule
    \multirow{2}{*}{\textbf{Operation}} & \multirow{2}{*}{\normalfont\textbf{Description}} & \multicolumn{2}{c}{\textbf{Execution Time (cycles)}} \\
                                        &                                                 & \multicolumn{2}{l}{\textbf{Analytical}\hspace{0.45in}\textbf{Meas.}} \\\midrule
    \tt{and\_16}        & 16-bit bit-wise and                         & $T_{\text{and}}$                        &    12 \\
    \tt{or\_16}         & 16-bit bit-wise or                          & $T_{\text{or}}$                         &    8 \\
    \tt{not\_16}        & 16-bit bit-wise not                         & $T_{\text{not}}$                        &    10 \\
    \tt{xor\_16}        & 16-bit bit-wise xor                         & $T_{\text{xor}}$                        &    12 \\
    \tt{ashift}         & int16 arithmetic shift                & $T_{\text{ash}}$                        &   15 \\
    \tt{add\_u16}       & uint16 element-wise addition                & $T_{\text{uadd}}$                       &   12 \\
    \tt{add\_s16}       & int16 element-wise addition                 & $T_{\text{sadd}}$                       &   13 \\
    % \tt{add\_gf16}      & gsi float16 element-wise addition           & $T_{\text{fadd}}$                       &  115 \\
    \tt{sub\_u16}       & uint16 element-wise subtraction             & $T_{\text{usub}}$                       &  15 \\
    \tt{sub\_s16}       & int16 element-wise subtraction              & $T_{\text{ssub}}$                       &  16 \\
    % \tt{sub\_gf16}      & gsi float16 element-wise subtraction        & $T_{\text{fsub}}$                       &  127 \\
    \tt{popcnt\_16}     & 16-bit population count                     & $T_{\text{popcnt}}$                     &    23 \\
    \tt{mul\_u16}       & uint16 element-wise multiplication          & $T_{\text{umul}}$                       &    115 \\
    \tt{mul\_s16}       & int16 element-wise multiplication           & $T_{\text{smul}}$                       &    201 \\
    \tt{mul\_f16}       & float16 element-wise multiplication         & $T_{\text{fmul}}$                       &     77 \\
    \tt{div\_u16}       & uint16 element-wise division                & $T_{\text{udiv}}$                       &     664 \\
    \tt{div\_s16}       & int16 element-wise division                 & $T_{\text{sdiv}}$                       &     739 \\
    % \tt{div\_imm\_gf16} & gsi float16 element-wise division by imm    & $T_{\text{fdiv}}$                       &     85 \\
    \hline
    \tt{eq\_16}         & 16-bit element-wise equal                   & $T_{\text{eq}}$                         &    13 \\
    \tt{gt\_u16}        & uint16 element-wise greater than            & $T_{\text{ugt}}$                        &    13 \\
    % \tt{gt\_gf16}       & gsi float16 element-wise greater than       & $T_{\text{fgt}}$                        &     63 \\
    \tt{lt\_u16}        & uint16 element-wise less than               & $T_{\text{ult}}$                        &    13 \\
    \tt{lt\_gf16}       & gsi float16 element-wise less than          & $T_{\text{flt}}$                        &    45 \\
    \tt{ge\_u16}        & uint16 greater than or equal                & $T_{\text{uge}}$                        &    13 \\
    \tt{le\_u16}        & uint16 less than or equal                   & $T_{\text{ule}}$                        &    13 \\
    \hline
    \tt{recip\_u16}     & uint16 element-wise reciprocal              & $T_{\text{recip}}$                      &    735 \\
    \tt{exp\_f16}       & float16 exponential                         & $T_{\text{exp}}$                        &   40295 \\
    \tt{sin\_fx}        & fixed-point sine                            & $T_{\text{sin}}$                        &    761 \\
    \tt{cos\_fx}        & fixed-point cosine                          & $T_{\text{cos}}$                        &    761 \\
    \hline
    \tt{count\_m}       & count marked entries                        & $T_{\text{cnt\_m}}$                     &   239 \\
    \tt{add\_subgrp\_s16} & int16 add sub groups in each group        & Eq.~\ref{eq:sg_add_equations}           & -- \\
  \bottomrule
  \end{tabular}}
  }
  % \vspace{0.05in}
  % \footnotesize
  % $b$ = element bit-width. $e$ is zero if $b =
  % 16$ and one otherwise. $C_1$ and $C_2$ indicate cycles due to control
  % processor instructions contained within our vector implementations;
  % $C_1$~= cycles that would vary with bit-width; $C_2$~= cycles that are
  % independent of bit-width. Rightmost column indicates average measured
  % number of cycles for 16-bit elements.
\end{table}

%% file: code/analytical_framework.tex
\begin{figure}[!htbp] 
    \centering
    \includegraphics[width=\linewidth]{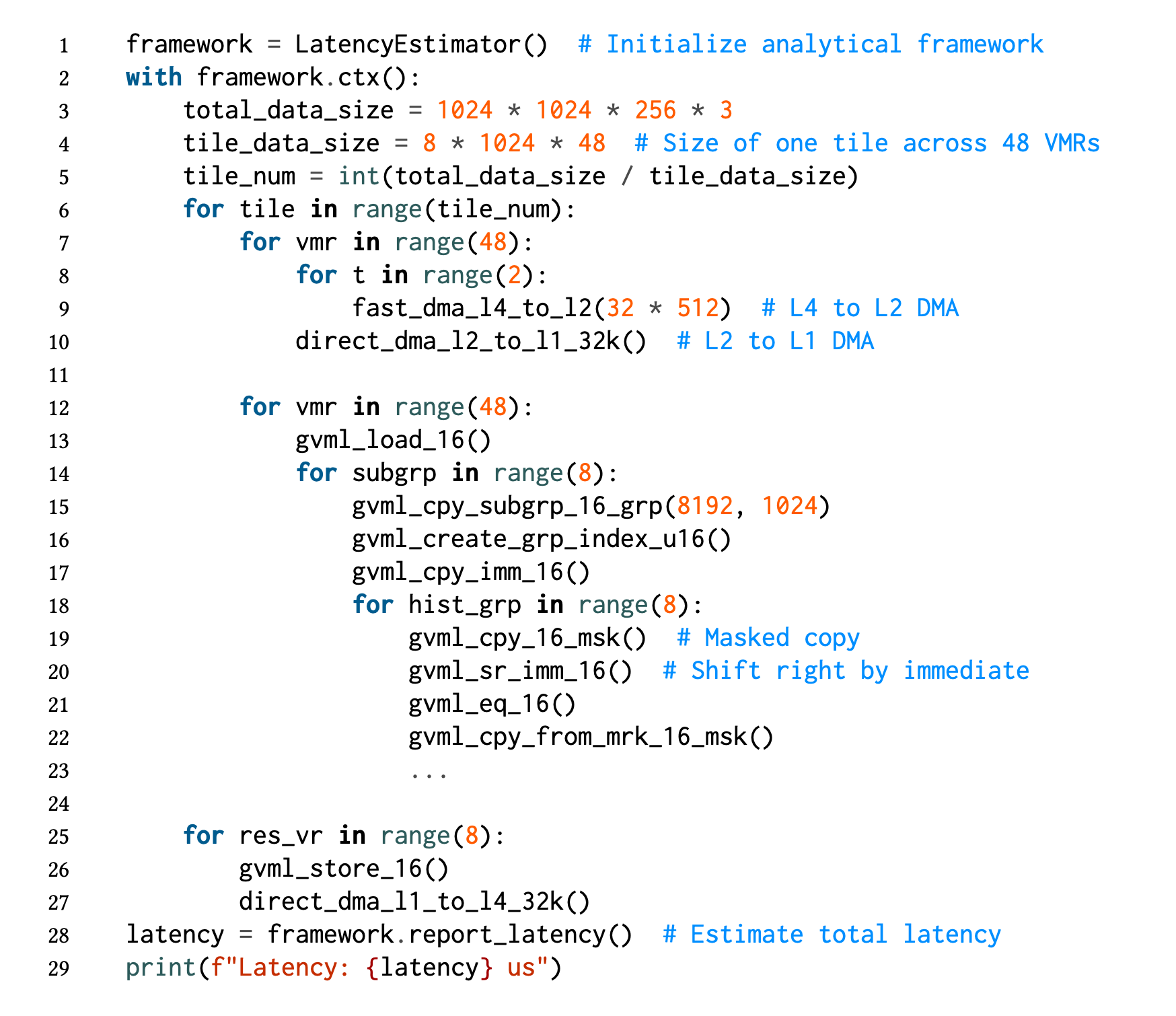}
  \caption{\revise{An example of modeling application latency with the analytical framework. We developed a Python function library with an interface similar to the GSI-provided C++ API. This example models the latency of the \texttt{Histogram} application from the Phoenix benchmark suite~\cite{ranger2007evaluating}.}}
  \label{fig:analytical_impl}
  \vspace{-.2in}
\end{figure}

%% file: sections/4-optimizations.tex
\section{Optimizing Realistic Workloads on Compute-in-SRAM}
\label{sec:optimizations}

Compute-in-SRAM provides substantial advantages in data parallelism and energy efficiency. However, it also introduces specific challenges and opportunities for optimization. Here, we use binary matrix multiplication as a motivating example to illustrate three key optimizations for compute-in-SRAM devices.

\subsection{Motivating Example}

Binary matrix multiplication is a crucial kernel for efficient machine learning, supporting workloads such as binary neural networks~\cite{zhang2021fracbnn, zhang2022pokebnn} and binarized transformers~\cite{wang2023bitnet, le2023binaryvit}. Compute-in-SRAM platform is a natural fit for this kernel due to its efficiency and speed in logical operations and integer addition. However, it is not easy to achieve high performance without careful consideration of data layout and data movement.

In the motivating example shown in Fig.~\ref{fig:matmul}, the input matrices $\mathbf{A}$ $(M, K)$ and $\mathbf{B}$ $(K, N)$ are bit-packed into \texttt{uint16} types along the $K$-axis. The binary matrix multiplication produces an output matrix $\mathbf{C}$ $(M, N)$ in \texttt{int16} type. The algorithm is depicted in Fig.~\ref{fig:matmul}(a). To implement this inner-product algorithm on an ultra-long vector processor, the baseline approach unrolls loop $j$, leading to the data layout illustrated in Fig.~\ref{fig:matmul}(c). We refer to this loop mapping scheme as spatial reduction vector mapping, as the reduction sum occurs spatially within the VR.
Let the VR length be $l=32768$. We consider the device DRAM as off-chip memory, assuming matrix $B$ fits in L1, the baseline operational intensity (OI) is:

\begin{equation}
    OI = \frac{M \cdot N \cdot K \cdot \alpha}{\left(MK \cdot \left\lfloor l/K \right\rfloor + KN + MN\right ) \cdot \tt{sf(u16)}}\,,
\end{equation}

\noindent where $\lfloor l/K\rfloor$ is the duplication factor of matrix $A$ due to loop $j$ unrolling, $\alpha$ is the number of logical and arithmetic operations on each scalar, and \texttt{sf()} denotes \texttt{size\_of()}. Matrix $A$ rows are duplicated in DRAM$\rightarrow$L2 and moved to L1, with a run-time cost of:

\begin{equation}
    T_A = \left(\frac{K \cdot \tt{sf(u16)}}{\text{BW}} + T_{\text{init}}\right) \cdot \left\lfloor \frac{l}{K} \right\rfloor \cdot M + M \cdot T_{\text{l2}\rightarrow \text{l1}}\,.
\end{equation}

\noindent We assume matrix $B$ is stored in a column-major layout in the device DRAM, and it fits in L1, then the run time cost of moving matrix $B$ is given by:

\begin{equation}
    T_B = \frac{N}{\lfloor l/K \rfloor} \cdot T_{\text{l4} \rightarrow \text{l1}}.
\end{equation}

\noindent For non-contiguous results in VR $C$, PIO transfers each element to L4, with a cost of:

\begin{equation}
    T_C = M \cdot N \cdot T_{\text{pio\_st}}
\end{equation}

\noindent The compute run-time cost is:

\begin{equation}
    T_{\text{MAC}} = \frac{N}{\lfloor l/K \rfloor} \cdot (T_{\text{xor}} + T_{\text{popcnt}} + T_{\text{ash}} + T_{\text{ssub}} + T_{\text{sg\_add}}(K, 1))
\end{equation}

\noindent and the total run-time cost is the sum of the data movement costs and the compute cost.

\subsection{Communication-Aware Reduction Mapping}

\begin{figure}[!htbp]
    \centering
    \includegraphics[width=\linewidth]{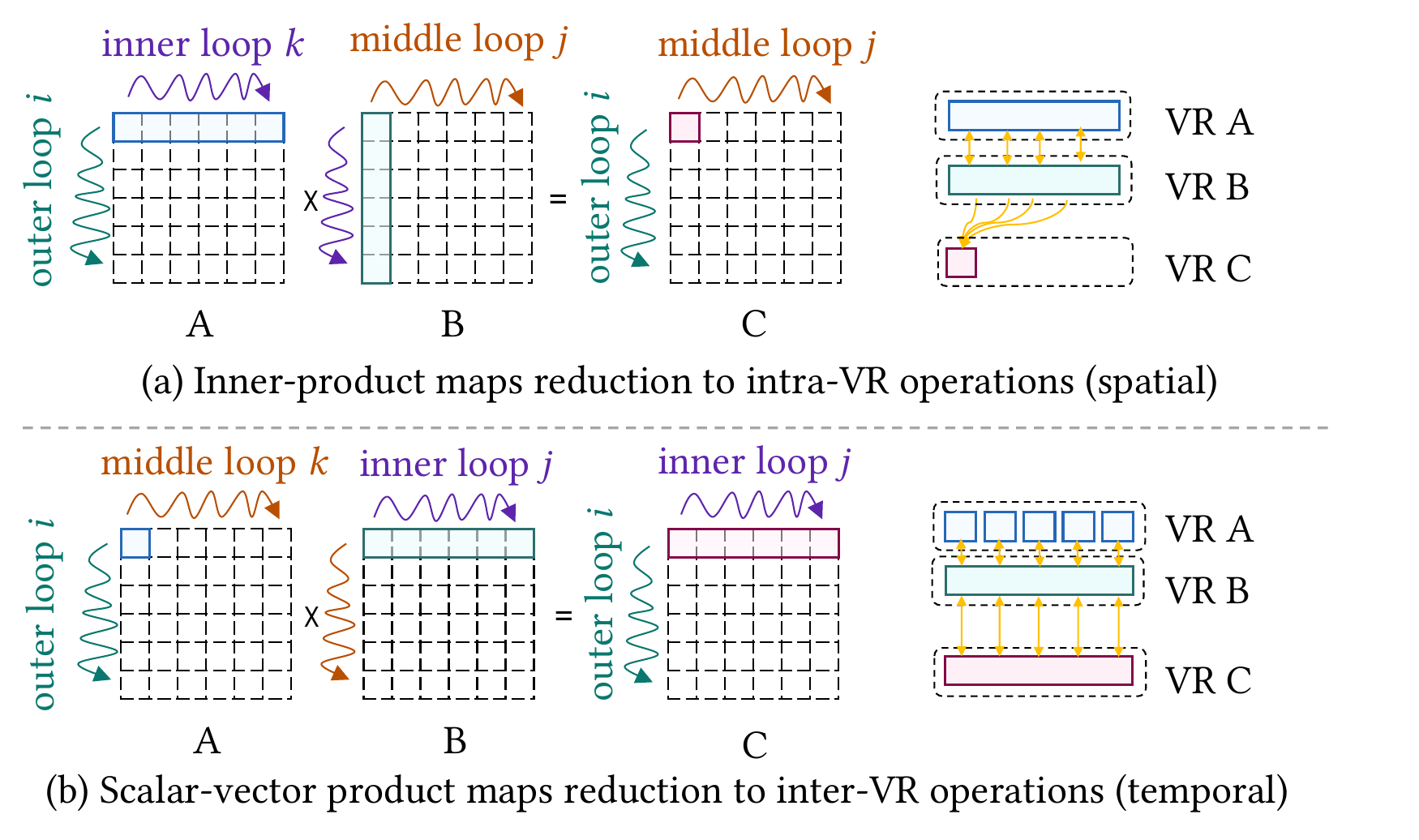}
    \vspace{-.3in}
    \caption{Reduction axis spatial vs. temporal mapping.}
    \label{fig:map}
\end{figure}

As outlined in the analytical framework: (1) intra-VR operations are more costly than inter-VR operations, and (2) using DMA to transfer the same amount of data is significantly cheaper than using PIO. 
Guided by these observations, we implement binary matrix multiplication as scalar-vector product (SVP)~\cite{de2020transformations}. As shown in Fig.~\ref{fig:map}, the reduction axis is mapped to the more efficient element-wise operations at each $k$ loop iteration. We refer to this loop mapping scheme as temporal reduction vector mapping. Additionally, the output data layout becomes contiguous, enabling fast DMA. Therefore, the compute run-time cost and matrix $\mathbf{C}$ movement cost reduce to:

\begin{equation}
    T_{\text{MAC}} = (T_{\text{xor}} + T_{\text{popcnt}} + T_{\text{ash}} + T_{\text{ssub}} + T_{\text{sadd}}) \cdot \frac{M}{\lfloor l/N \rfloor} \cdot K,
\end{equation}

\begin{equation}
    T_C = \frac{M}{\lfloor l/N \rfloor} \cdot T_{\text{l4} \rightarrow \text{l1}}.
\end{equation}

\begin{figure}[!tbp]
    \centering
    \includegraphics[width=0.45\textwidth]{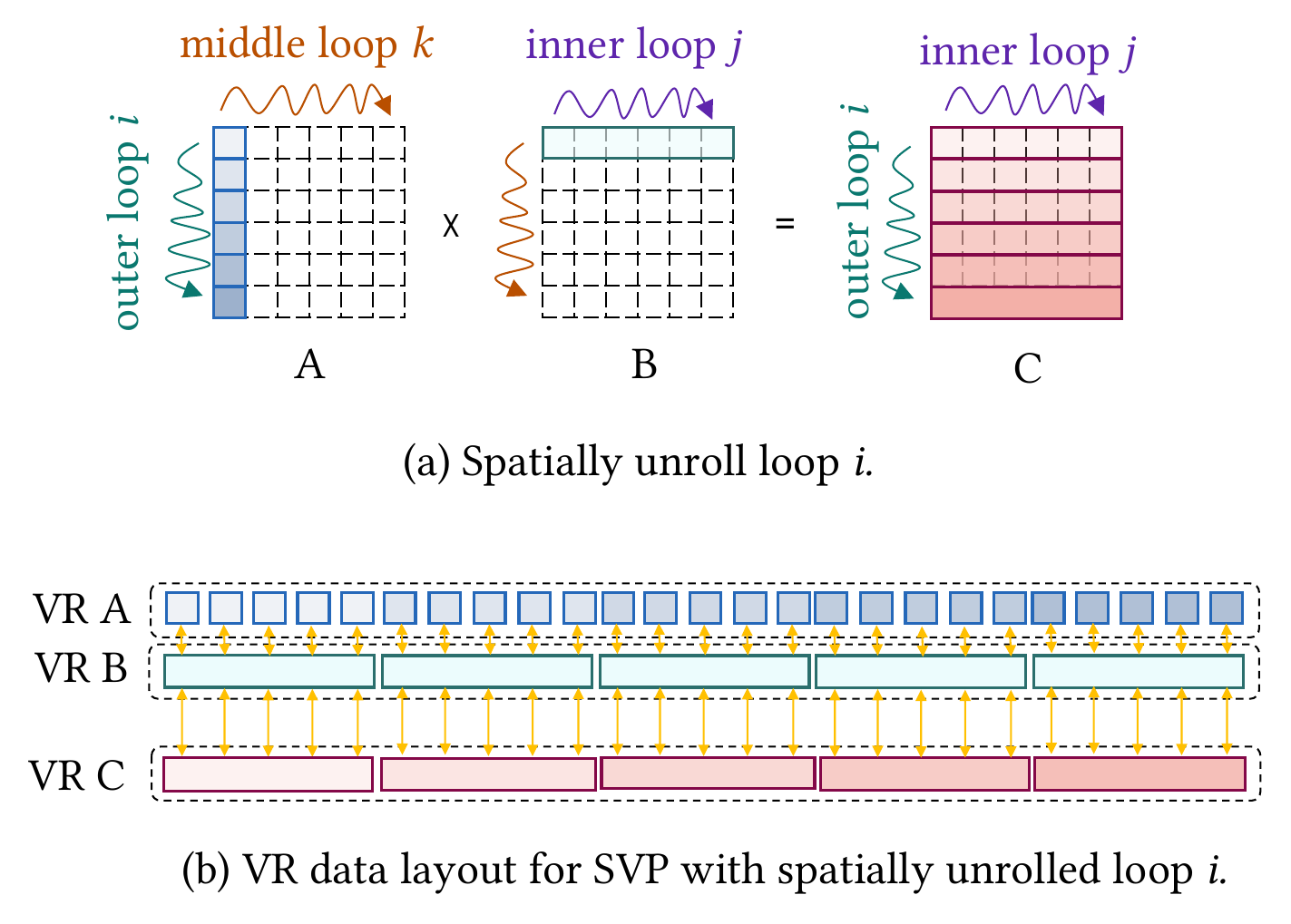}
    \vspace{-.1in}
    \caption{Spatially unrolling $i$-axis fully utilizes the VR, enables inter-VR reduction, and achieves a contiguous layout of results.}
    \label{fig:unroll}
\end{figure}

Since all bit processors operate in parallel, higher VR occupancy translates to improved computational efficiency. To achieve this, we unroll loop \(i\) to fully utilize the VR, as shown in Fig.~\ref{fig:unroll}(a). Loop \(i\) is specifically chosen for unrolling to maintain the temporal mapping of loop \(k\). Consequently, this approach results in two levels of data duplication in the VR layout, as shown in Fig.~\ref{fig:unroll}(b): the scalars from \(A\) are duplicated due to the spatial unrolling of loop \(j\), and rows from \(B\) are duplicated due to the spatial unrolling of loop \(i\). This data layout enables opportunities for data reuse and memory access coalescing. We implement the scalar duplication of matrix $A$ as a lookup operation from L3. Therefore, the OI for the scalar-vector product becomes

\begin{equation}
    OI = \frac{M \cdot N \cdot K \cdot \alpha}{(MK + NK \cdot \lfloor l / N \rfloor + MN) \cdot \tt{sf(u16)}}
\end{equation}

\noindent Assuming matrix $A$ is stored in row-major order, the run-time cost of moving it involves transferring from L4 to L3, followed by duplication via lookup:

\begin{equation}
    T_A = M \cdot K / \text{BW} + T_{\text{init}} + T_{\text{lookup}}(N \cdot K) \cdot \frac{M}{\lfloor l/N \rfloor} \cdot K
\end{equation}

\noindent For B, loop $i$ spatial unrolling incurs duplication of factor $\lfloor l/N \rfloor$, the run-time cost of moving matrix $B$ becomes
\begin{equation}
    T_B = \left( \frac{N \cdot \tt{sf(u16)}}{\text{BW}} + T_{\text{init}} \right) \cdot \left\lfloor \frac{l}{N} \right\rfloor \cdot K + K \cdot T_{\text{l2} \rightarrow \text{l1}}
\end{equation}

\subsection{DMA Coalescing}

\begin{figure}[!tbp]
    \centering
    \includegraphics[width=\linewidth]{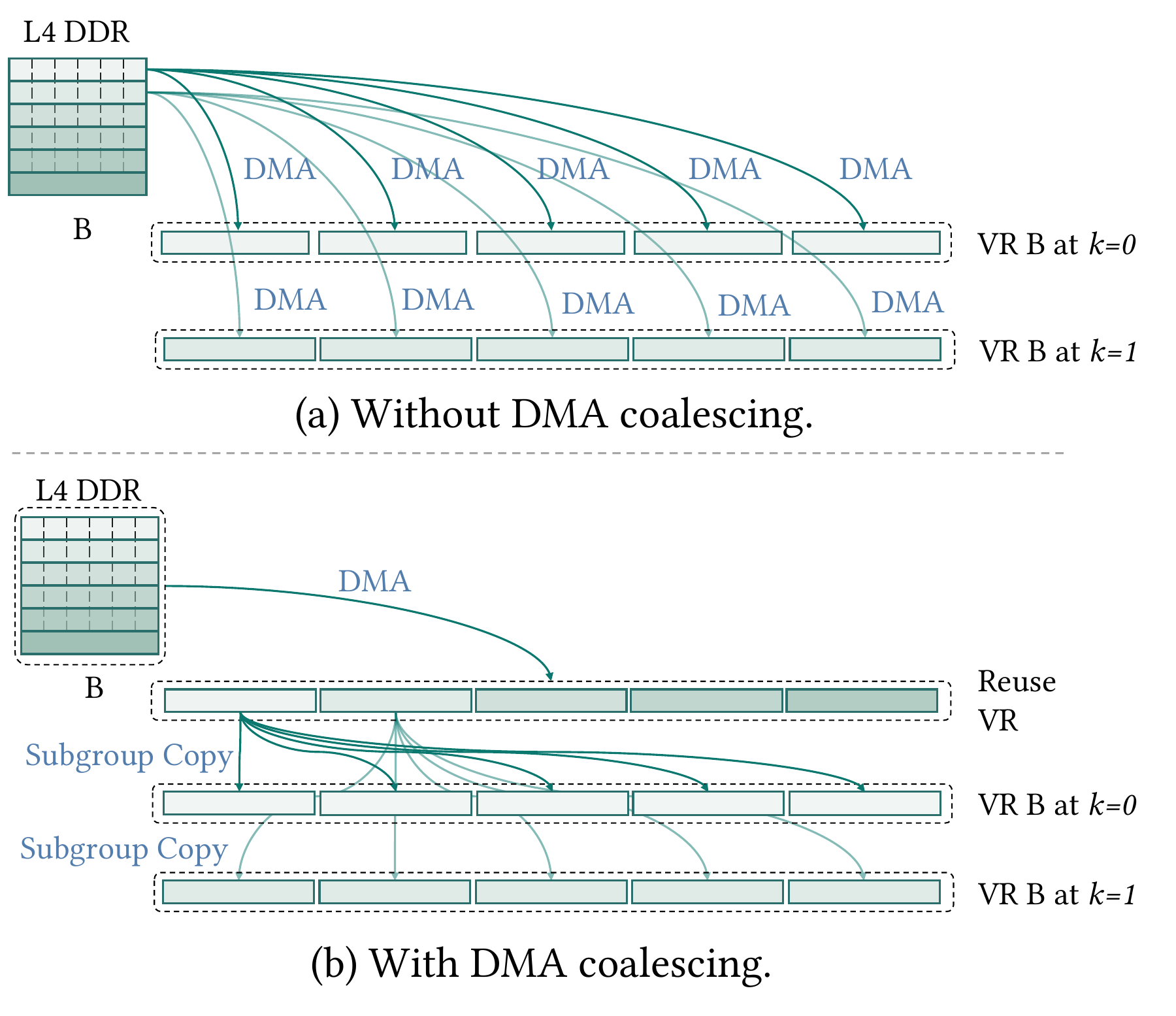}
    \vspace{-.15in}
    \caption{Coalescing DMA -- leverage subgroup copy to remove redundant data access and increase data reuse.}
    \label{fig:dma}
\end{figure}

Once we optimize the data layout, a new bottleneck of data duplication emerges. As seen in Fig.~\ref{fig:unroll}, one form of data duplication is that of duplicating a chunk of data across an entire VR.
In Fig.~\ref{fig:dma}(a), we see that DMA transactions can be used for data duplication. However, this approach is bandwidth-inefficient since accessing off-chip memory incurs high latency, and multiple DMA transactions add initiation overhead. Because the same chunk of data from \(B\) is accessed in each iteration of loop \(k\), we can coalesce these DMA accesses to avoid redundant data movement. Specifically, we combine DMA transactions on multiple rows of \(B\) into a single transaction, minimizing initiation overhead. 

To implement this, we introduce a reuse VR to store the initial DMA result. Using the subgroup copy capability, each row of \(B\) is arranged in a subgroup and copied to fill the VR at each iteration of loop \(k\). Notably, subgroup copy can also target a portion of the VR, providing flexibility when duplicating only part of the data.
This optimization results in a lower run-time cost of moving matrix $B$:

\begin{equation}
    T_B = \left\lceil \frac{K \cdot N}{l} \right\rceil \cdot T_{\text{l4} \rightarrow \text{l1}} + K \cdot T_{\text{cpy\_sgp}}
\end{equation}

\noindent Since DMA coalescing also removes duplicate data movement from L4, the OI is also improved:

\begin{equation}
    OI = \frac{M \cdot N \cdot K \cdot \alpha}{(MK + NK + MN) \cdot \tt{sf(u16)}}.
\end{equation}

\subsection{Broadcast-Friendly Data Layout}

\begin{figure*}[!htbp]
    \centering
    \begin{tabular}{cc}
        \includegraphics[width=0.51\textwidth]{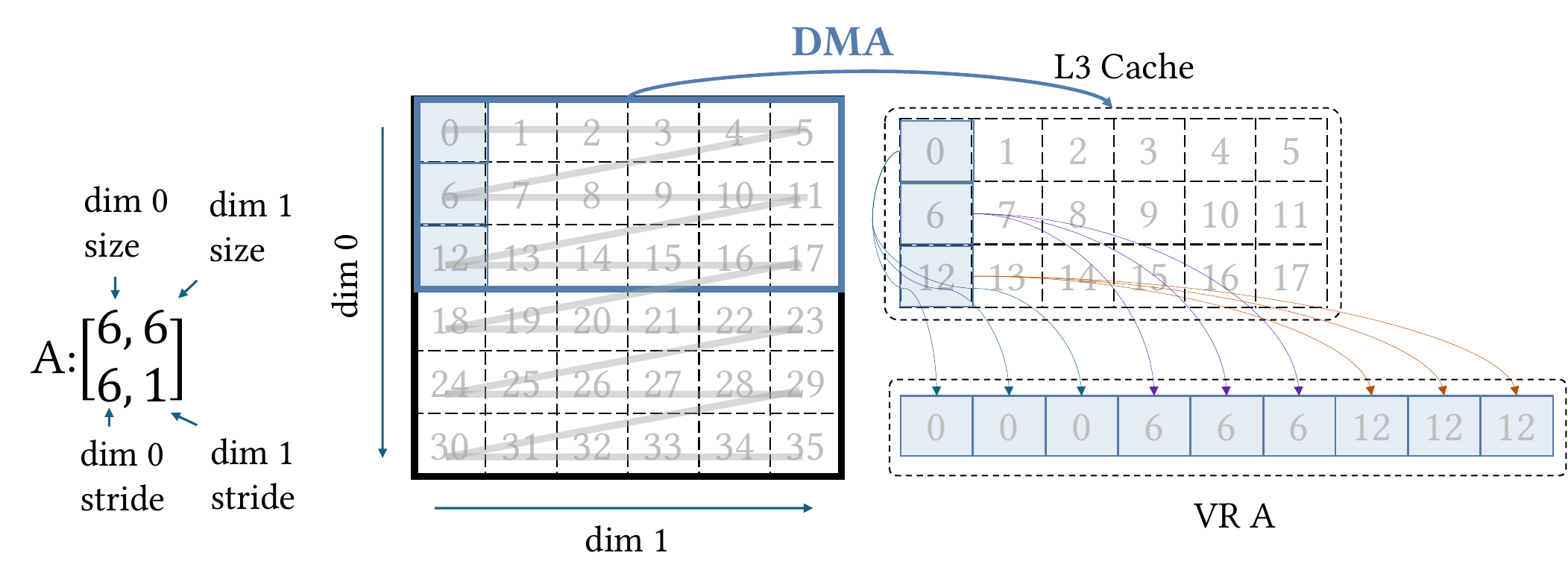} &
        \includegraphics[width=0.43\textwidth]{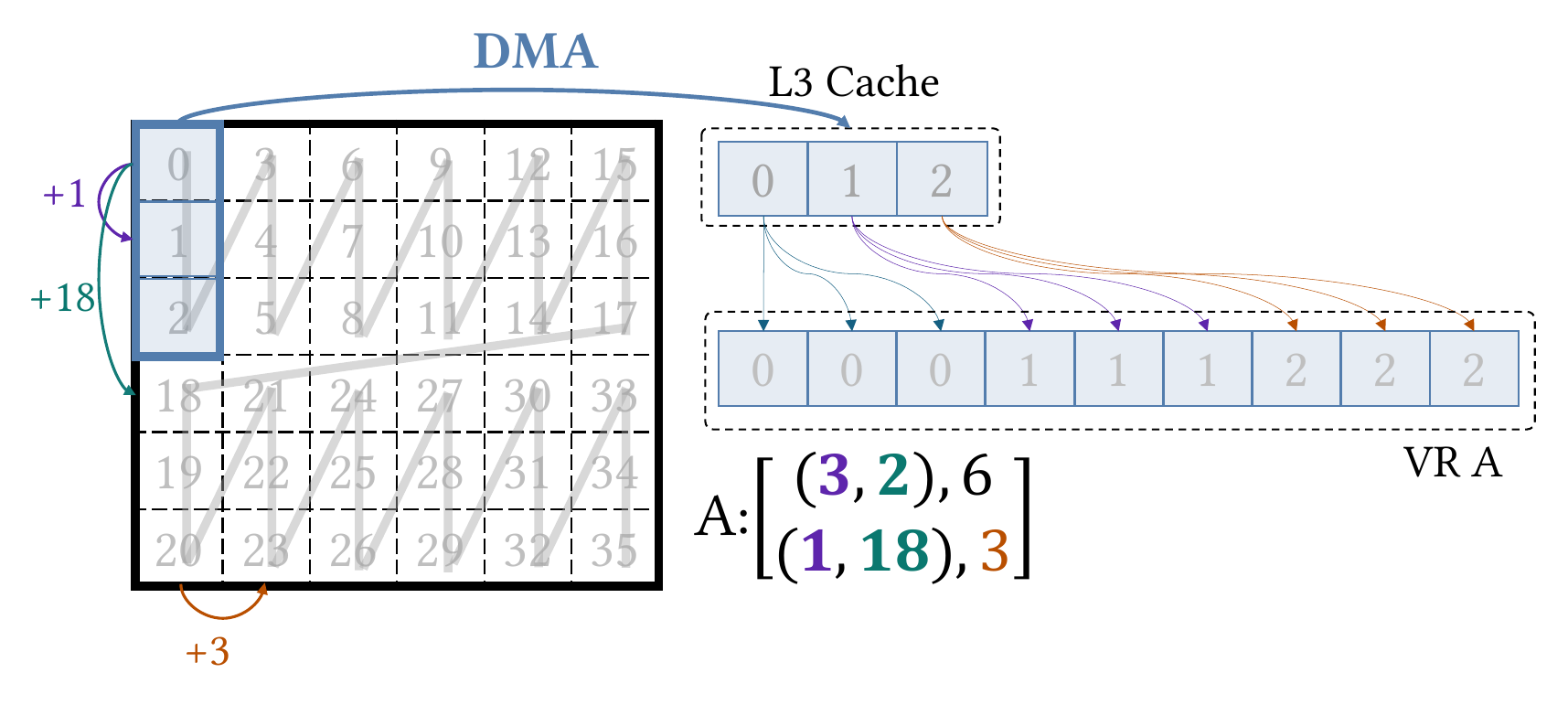} \\
        (a) Row-major data layout. &
        (b) A broadcast-friendly data layout \\
    \end{tabular}
    \caption{\textbf{Broadcast-friendly data layout example} -- (a) a row-major data layout requires a lookup table of size 18. (b) A broadcast-friendly data layout only requires a lookup table size of 3.}
    \label{fig:broadcast}
\end{figure*}

After removing the redundant DMA operations, the bottleneck shifts to the lookup operation used to broadcast scalars in $\mathbf{A}$. As shown in Table~\ref{tab:analytical-memory}, the lookup latency increases with the size of the lookup table, prompting us to reduce its size. 
Fig.~\ref{fig:broadcast} illustrates the lookup operation, where three scalars are broadcast each time, highlighted by the \textcolor{blue}{blue}-filled boxes. In the row-major layout shown in Fig.~\ref{fig:broadcast}(a), the broadcast window initially covers indices 0, 6, and 12, and then \revise{moves} to indices 1, 7, and 13 in the next iteration.
Since the lookup table must be a contiguous chunk of memory, the lookup table size is at least 18 to broadcast the first three rows.
To reduce the lookup table size, we change the data layout to a broadcast-friendly format, shown in Fig.~\ref{fig:broadcast}(b). The broadcast window initially covers indices 0, 1, 2, and then moves on to 3, 4, 5. Therefore, the lookup table sizes can be reduced to 3.
We express this data layout as dimension sizes and strides, where decomposed sizes and strides are expressed as tuples, as shown in Fig.~\ref{fig:broadcast}. This format is proposed by Graphene~\cite{hagedorn2023graphene}.
For the motivating example, this optimization reduces the lookup table size for broadcasting matrix $\mathbf{A}$ from $K\cdot N$ to $N$, thereby reducing the cost of data movement to:

\begin{equation}
    T_A = M \cdot K / \text{BW} + T_{\text{init}} + T_{\text{lookup}}(N) \cdot \frac{M}{\lfloor l/N \rfloor} \cdot K.
\end{equation}

In summary, we demonstrate how these three key optimizations for data layout and movement reduce both input/output transfer costs and computation costs for compute-in-SRAM devices. 

%% file: sections/5-evaluation.tex
\section{Evaluation}

Using the GSI APU as a commercial example, this section validates the analytical framework and evaluates the real-world performance of compute-in-SRAM with the proposed optimizations. First, a latency breakdown of binary matrix multiplication highlights the individual contributions of each optimization. Next, a benchmark study validates the analytical framework and identifies workload characteristics well-suited for in-SRAM computing. Finally, an end-to-end retrieval-augmented generation study on large corpora compares the performance and energy efficiency of compute-in-SRAM against CPU and GPU platforms.

We use the GSI Leda-E APU (500 MHz clock frequency), an Intel Xeon Gold 6230R CPU (2.1 GHz, 1.6 MB L1 cache, 52 MB L2 cache, 71.5 MB L3 cache), and an NVIDIA A6000 GPU for comparison. Latency measurements on the GSI APU are obtained using control processor cycle counts. Energy profiling is performed using a Texas Instruments UCD9090 voltage monitor and Renesas ISL8273M power modules on board, which provide point-of-load regulation and current telemetry.

\subsection{Binary Matrix Multiplication}
\revise{We use a $1024\times 1024$ binary matrix multiplication kernel as a microbenchmark to analyze and demonstrate the impact of the proposed optimizations.}

\begin{figure}[!htbp]
    \centering
    \includegraphics[width=.9\columnwidth]{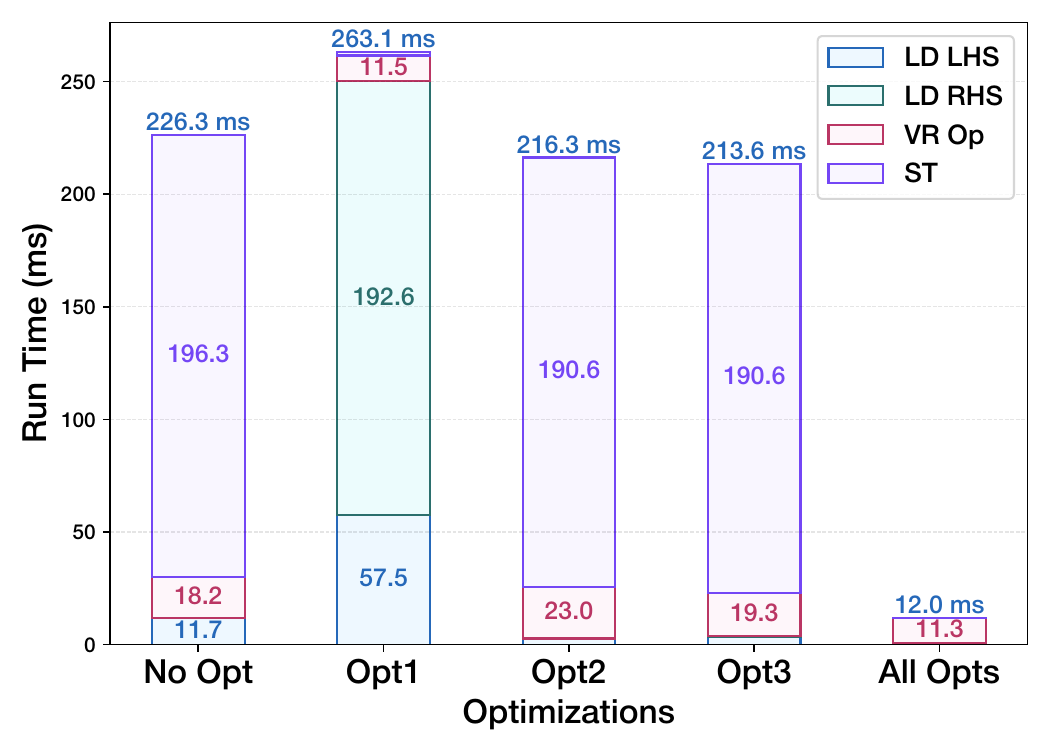}
    \caption{\revise{Binary matrix multiplication runtime breakdown with different optimizations. Opt1: communication-aware reduction mapping. Opt2: DMA Coalescing. Opt3: broadcast-friendly data layout.}}
    \label{fig:bmatmul}
    \vspace{-.1in}
\end{figure}

Fig.~\ref{fig:bmatmul} illustrates the latency breakdown from the baseline implementation to the optimized versions. \revise{Key} operations include: LD LHS / RHS, loading matrices from off-chip memory to L1 via DMA, PIO, or lookup; VR Ops, on-chip operations like subgroup copies and computations; and ST, storing results back to off-chip memory.

\revise{
\noindent We use an inner-product algorithm as the baseline implementation (described in Section~\ref{sec:data_movement}), which is bottlenecked by result data movement due to costly PIO stores for non-contiguous outputs. Applying communication-aware reduction mapping (opt1) reduces this overhead by enabling efficient DMA transfers, though it increases RHS matrix loading time due to data duplication. Adding DMA coalescing (opt2) further improves LHS loading by replacing PIO with faster DMA, at the cost of additional vector register operations for subgroup copies. Introducing a broadcast-friendly data layout (opt3) also accelerates LHS loading, but the overall bottleneck remains in the result write back. 
When all three optimizations are combined, results become contiguous and can be transferred using DMA. We also apply DMA coalescing for the RHS matrix using $k$-axis data packing and adopt a broadcast-friendly format 
}
$
\bigl[\begin{smallmatrix}
(32,\ 32) & 64 \\ 
(1,\ 2048) & 32 
\end{smallmatrix}\bigr]
$
\revise{
for the LHS matrix broadcasting, which yields an end-to-end latency of 12.0\,ms, an 18.9$\times$ improvement over the baseline latency of 226.3\,ms.}

\revise{
This result demonstrates that while individual optimizations may yield modest speedups, they enable opportunities for further improvements. For example, communication-aware reduction mapping enables DMA coalescing to RHS matrix loading and facilitates the use of a broadcast-friendly layout for the LHS matrix. Ultimately, combining all optimizations leads to substantial performance gains.
}

\begin{figure*}[!htbp]
    \centering
    \includegraphics[width=\textwidth]{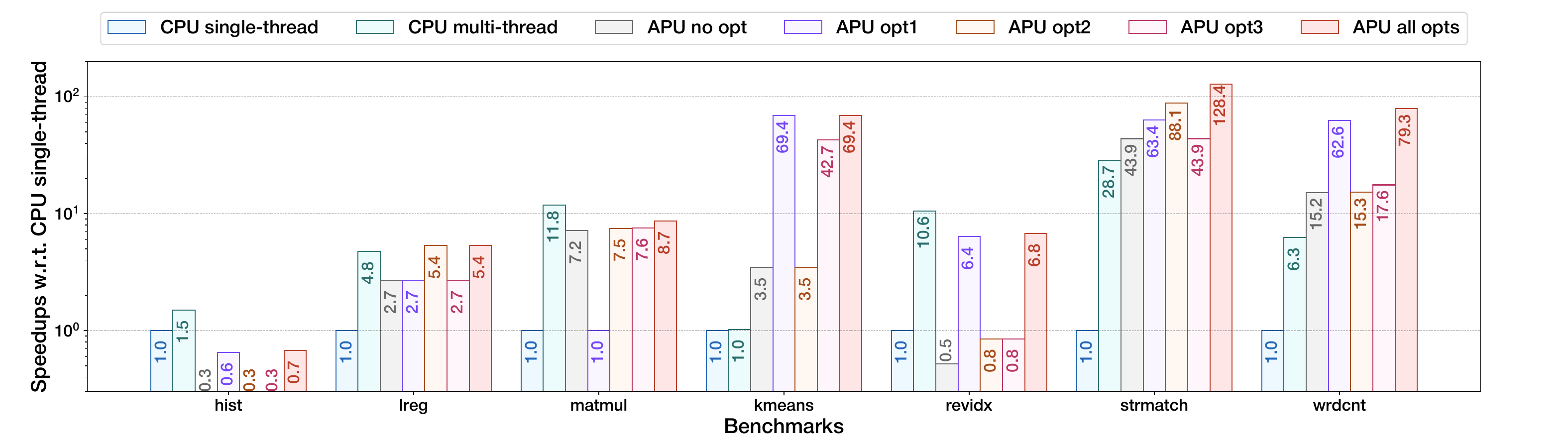}
    \vspace{-.3in}
    \caption{\revise{Latency comparison across workloads from the \textit{Phoenix Benchmark Suite}, normalized to the single-threaded Intel Xeon Gold CPU baseline. Opt1: communication-aware reduction mapping. Opt2: DMA Coalescing. Opt3: broadcast-friendly data layout.}}
    \label{fig:phoenix}
\end{figure*}

\subsection{Phoenix Benchmarks}
\label{sec:phoenix}

We evaluate the optimized GSI APU performance using the Phoenix Benchmark suite~\cite{ranger2007evaluating}, comparing it against two baselines: single-threaded and multi-threaded CPU implementations.
We select Phoenix for its data-intensive benchmarks and its prior use in compute-in-SRAM studies such as CAPE~\cite{caminal2021cape}.
The official Phoenix benchmark provides optimized CPU implementations, either single-threaded or multi-threaded. We use the official repository\footnote{\url{https://github.com/kozyraki/phoenix}}, with the multi-threaded version configured for up to 16 threads using the MapReduce programming model. Table~\ref{tab:phoenix} summarizes the eight applications in the suite, including their CPU instruction count (measured with Valgrind) and APU $\mu$Code instruction count (reported by the Vector Command Unit). Input sizes range from 10 MB to 1.5 GB. 
We measure the total APU kernel latency, covering data movement from device memory to L1 and back.
We achieve optimized performance on these benchmarks by applying all proposed data movement and data layout optimizations.
% \zz{are we applying all the proposed optimizations on each benchmark? add a line to clarify}

\input{tables/phoenix}

\revise{
Figure~\ref{fig:phoenix} compares the latency of the APU implementations against CPU baselines. Relative to the single-threaded CPU, the APU implementation with all optimizations applied achieves an average speedup of 41.8$\times$ (mean), 14.4$\times$ (geometric mean), and a peak speedup of 128.3$\times$. Compared to the multi-threaded CPU execution, the APU achieves an average speedup of 12.5$\times$ (mean), 2.6$\times$ (geometric mean), and a maximum speedup of 68.1$\times$.
}

\subsubsection{Results Analysis}

\revise{
The APU implementations shown in Fig.~\ref{fig:phoenix} include a baseline with no optimizations, as well as versions applying only communication-aware reduction mapping (Opt1), only DMA coalescing (Opt2), only broadcast-friendly data layout (Opt3), and all three optimizations together (APU all opts). Individually, communication-aware reduction mapping provides large gains in workloads involving comparison or distance computation over large volumes of data, such as \texttt{kmeans}, \texttt{reverse index}, \texttt{string match}, and \texttt{word count}. DMA coalescing reduces data movement costs in cases where data duplication is required (e.g., \texttt{matmul}) or where input data can be packed to improve vector register utilization (e.g., \texttt{linear regression}, \texttt{string match}). Broadcast-friendly data layout is beneficial when scalar values are broadcast via lookup operations, such as in \texttt{kmeans}, although these opportunities often emerge only after other optimizations have been applied. Overall, we observe that applying all three optimizations consistently yields greater performance improvements than applying any single optimization in isolation.
}

The fully optimized results on the benchmark suite suggest that compute-in-SRAM platforms are best suited for a specific subset of applications. The optimized APU implementation outperforms a multi-threaded CPU on linear regression, k-means, string match, and word count: applications characterized by high data parallelism and minimal intra-VR computation. With the proposed optimizations, most arithmetic operations are efficiently mapped to inter-VR element-wise instructions, and data duplication overhead is reduced. In contrast, other applications including histogram, matrix multiply, reverse index, still involve frequent intra-VR operations and fine-grained element access due to their algorithmic nature, limiting the performance benefits from compute-in-SRAM acceleration.

% \begin{figure}[!htbp]
%     \centering
%     \includegraphics[width=\columnwidth]{figs/runtime_breakdown.pdf}
%     \caption{Phoenix benchmarks latency breakdown.}
%     \label{fig:phoenix-breakdown}
% \end{figure}

\subsubsection{Analytical framework validation}
\label{sec:validate}

\input{tables/validation}

We validate the analytical framework using the Phoenix Benchmark suite by comparing the measured latency with the predicted latency. Table~\ref{tab:validation} summarizes the results across the eight benchmarks. On average, the analytical framework achieves 97.3\% accuracy, with a maximum error of 6.2\%. The primary source of error arises from the model's inability to account for memory subsystem details or cache behavior.

% \subsubsection{Comparison with CAPE}

% Although a direct comparison with CAPE on Phoenix is not feasible due to the lack of absolute latency values and the different CPU baseline used for normalization, we discuss the similarities and differences in the results. CAPE achieves an average 14$\times$ speedup on the Phoenix benchmarks compared to a two-core area-equivalent CPU~\cite{caminal2021cape}.
% The APU shows comparable high speedups for string match, linear regression, and k-means, where most operations are mapped to inter-VR computations. The APU achieves lower speedups on benchmarks dominated by inter-VR operations.
% Two key architectural and system-level differences explain this disparity: (1) the APU lacks the pipelined reduction tree found in CAPE, which optimizes intra-VR reductions, and (2) CAPE is equipped with high-bandwidth memory (HBM) providing 128 GB/s bandwidth, while the APU uses DDR4 memory with only 23.8 GB/s bandwidth. These limitations significantly impact the APU's performance on memory-intensive applications.

\subsection{Retrieval-Augmented Generation (RAG)}
\label{sec:rag}

% Retrieval-Augmented Generation (RAG) improves language model response quality by retrieving relevant information from external knowledge bases during generation~\cite{lewis2020retrieval, gao2023retrieval}. This typically involves embedding queries and documents, followed by a vector similarity search for the top-k results~\cite{salemi2024evaluating}. Handling vast corpora often forces CPU/GPU implementations to use Approximate Nearest Neighbor Search (ANNS), trading accuracy for lower latency and memory usage. Unfortunately, this approximation can degrade generation accuracy substantially (22\%-53\% loss with \texttt{Llama-8B} and \texttt{Llama-80B} on representative RAG applications, reported by~\cite{quinn2025rag}) compared to performing an Exact Nearest Neighbor Search (ENNS). Compute-in-SRAM platforms offer a compelling alternative: their inherent massive data parallelism makes them well-suited to accelerate ENNS efficiently, potentially avoiding the accuracy compromise. This section investigates the benefits of using an optimized compute-in-SRAM implementation for ENNS acceleration within RAG, focusing on end-to-end latency reduction and energy efficiency gains.
Retrieval-Augmented Generation (RAG) improves language model responses by retrieving relevant knowledge during generation~\cite{lewis2020retrieval, gao2023retrieval}. It embeds queries and documents, then performs vector similarity search~\cite{salemi2024evaluating}. Large corpora often require CPUs/GPUs to use Approximate Nearest Neighbor Search (ANNS), trading accuracy for latency and memory. This can cause significant accuracy loss (22\%–53\% for \texttt{Llama-8B} and \texttt{Llama-80B}~\cite{quinn2025rag}) compared to Exact Nearest Neighbor Search (ENNS). Compute-in-SRAM platforms, with massive parallelism, can accelerate ENNS efficiently, avoiding this compromise. We study optimized compute-in-SRAM ENNS for RAG, focusing on latency and energy benefits.

\subsubsection{Experimental Setup}

% system setup
We implement the ENNS RAG retrieval process on the GSI APU. However, the device's limited DDR bandwidth (23.8 GB/s) would unequivocally create an off-chip memory bottleneck, hindering a fair performance comparison. To mitigate this, we model a more representative off-chip memory system by simulating HBM2e memory (16 GB, 2 ranks, 8 channels, 1.6 GHz, yielding 380–420 GB/s peak bandwidth) using Ramulator 2~\cite{luo2023ramulator} and \revise{DRAMPower 5.0~\cite{drampower}}. The compute-in-SRAM performance results presented incorporate these simulated off-chip memory timings, while all other components—including on-chip data movement, computation, and system overheads are measured directly on the GSI APU hardware.

% corpus, rag setup
We use \texttt{Llama3.1-8B}~\cite{grattafiori2024llama} with 16-bit number format as the generation model and sample questions from the Natural Questions (NQ) dataset~\cite{kwiatkowski2019natural}. The evaluation system comprises two GPUs (one dedicated to generation and the other to retrieval), a CPU, and a GSI APU. Generation runs on a single GPU, and retrieval is performed using ENNS across three corpus sizes: 10\,GB, 50\,GB, and 200\,GB, on CPU, GPU, and a compute-in-SRAM accelerator. Each corpus is chunked into segments of 16{,}384 tokens. \revise{As a result, the 10\,GB corpus contains 163K chunks (120\,MB embedding size), the 50\,GB corpus contains 819K chunks (600\,MB embedding size), and the 200\,GB corpus contains 3.3M chunks (2.4\,GB embedding size).}

% benchmark methodology
\revise{For both the GPU and compute-in-SRAM accelerator, corpus embeddings are transferred to device memory once at the start of the workload. All subsequent queries are served without reloading the embeddings. Meanwhile, the corpus chunks reside in the CPU's main memory.}
In the results that follow, we report the time-to-interactive latency on each platform—also referred to as time-to-first-token latency—which serves as the primary metric for evaluating the interactivity of the LLM inference system. All latency results are averaged across 10 queries.

\subsubsection{Software Configurations}
We evaluate RAG performance on the CPU and GPU using FAISS~\cite{douze2024faiss}, a widely adopted library for efficient similarity search and clustering of dense vectors at scale. Our experiments use FAISS v1.7.2 to run ENNS inner product search with \texttt{IndexFlat}, leveraging AVX512 intrinsics and OpenMP-based multithreading on the CPU.

\subsubsection{End-to-End RAG Performance}

\begin{figure}[!htbp]
    \centering
    \includegraphics[width=\columnwidth]{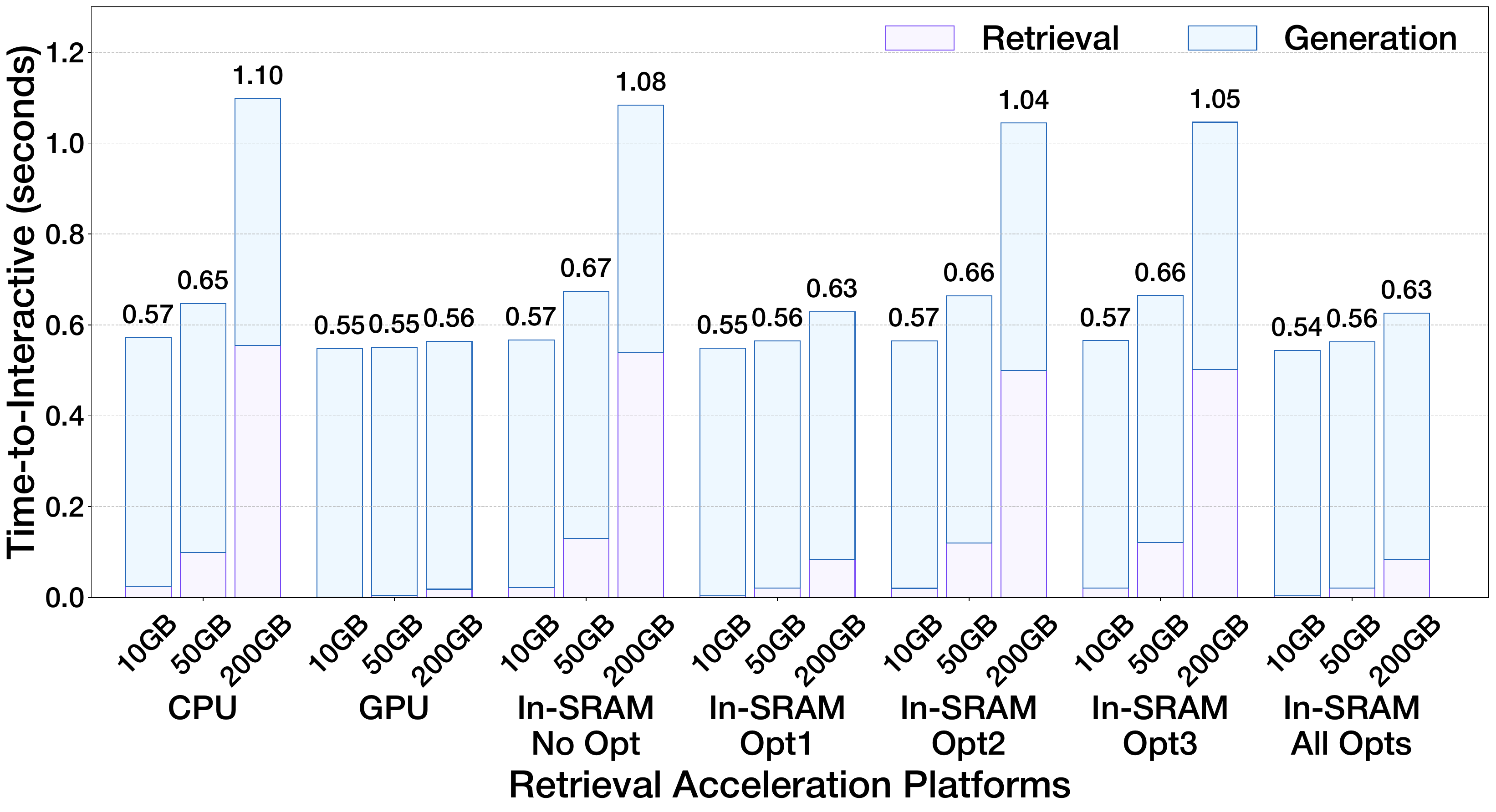}
    \caption{\revise{Inference time breakdown of CPU, GPU, vs. compute-in-SRAM with and without optimizations. The \texttt{Llama3.1-8B} generative model runs on a dedicated GPU. Opt1: communication-aware reduction mapping. Opt2: DMA Coalescing. Opt3: broadcast-friendly data layout.}}
    \label{fig:rag}
\end{figure}

% As shown in Fig.~\ref{fig:rag}, retrieval accounts for an increasing portion of end-to-end inference time as corpus size scales. For CPU-based retrieval, it grows from just 4.3\% of total latency at 10~GB to 50.5\% at 200~GB, making retrieval the dominant bottleneck. To address this, we implement a communication-aware reduction mapping and broadcast-friendly query layout on the compute-in-SRAM accelerator, optimizing inner-product search by mapping reduction operations to inter-VR instructions. This yields retrieval speedups of 7.5$\times$, 5.4$\times$, and 7.3$\times$ over CPU at 10~GB, 50~GB, and 200~GB respectively. These improvements translate to 1.05$\times$, 1.15$\times$, and 1.78$\times$ end-to-end speedups, demonstrating the growing impact of retrieval-side acceleration. Compared to the unoptimized compute-in-SRAM baseline, the proposed optimization reduces retrieval latency by up to 6.5$\times$. Notably, the optimized compute-in-SRAM system matches GPU-level end-to-end latency, underscoring the effectiveness of the proposed optimizations.

As shown in Fig.~\ref{fig:rag}, retrieval accounts for an increasing portion of end-to-end inference time as corpus size scales (CPU-based retrieval: 4.3\% at 10~GB $\rightarrow$ 50.5\% at 200~GB). We optimize inner-product search on the compute-in-SRAM accelerator via communication-aware reduction mapping and a broadcast-friendly query layout, mapping reductions to inter-VR instructions. Retrieval speedups over CPU are 6.3$\times$/4.8$\times$/6.6$\times$ at 10/50/200~GB, yielding 1.05$\times$/1.15$\times$/1.75$\times$ end-to-end gains. Versus an unoptimized compute-in-SRAM baseline, retrieval latency reduces up to 6.4$\times$. The optimized system attains GPU-level end-to-end latency, underscoring the effectiveness of retrieval-side acceleration.

\subsubsection{Optimization Impact and Retrieval Latency Breakdown}
APU retrieval latency for RAG at 10/50/200~GB is 21.8/129.5/539.2\,ms without optimization, comparable to CPU performance but slower than GPU. Communication-aware reduction mapping (opt1) addresses output data movement bottlenecks, cutting retrieval latency to 4.0/21.0/86.1\,ms. DMA coalescing (opt2) and a broadcast-friendly layout (opt3) give modest standalone gains but compound with opt1; with all three, latency drops to 3.9/20.6/84.2\,ms. Figure~\ref{fig:rag} shows that while opt2 and opt3 have limited standalone effect, they enhance opt1 by improving data movement and vector utilization.

As shown in Table~\ref{tab:rag-breakdown}, most of the optimized compute-in-SRAM retrieval speedup comes from the distance calculation stage. The key is communication-aware reduction mapping—which maps inner-product reductions to inter-VR ops to reduce intra-VR movement and improves alignment (e.g., embedding-load time drops from 8.2$\,$ms to 6.1$\,$ms at 200~GB). A broadcast-friendly layout further lowers query-broadcast overhead and boosts vector reuse, adding additional compute-time savings.

\subsubsection{Energy Efficiency Comparison with GPU}
\begin{figure}[t]
    \centering
    \includegraphics[width=.75\columnwidth]{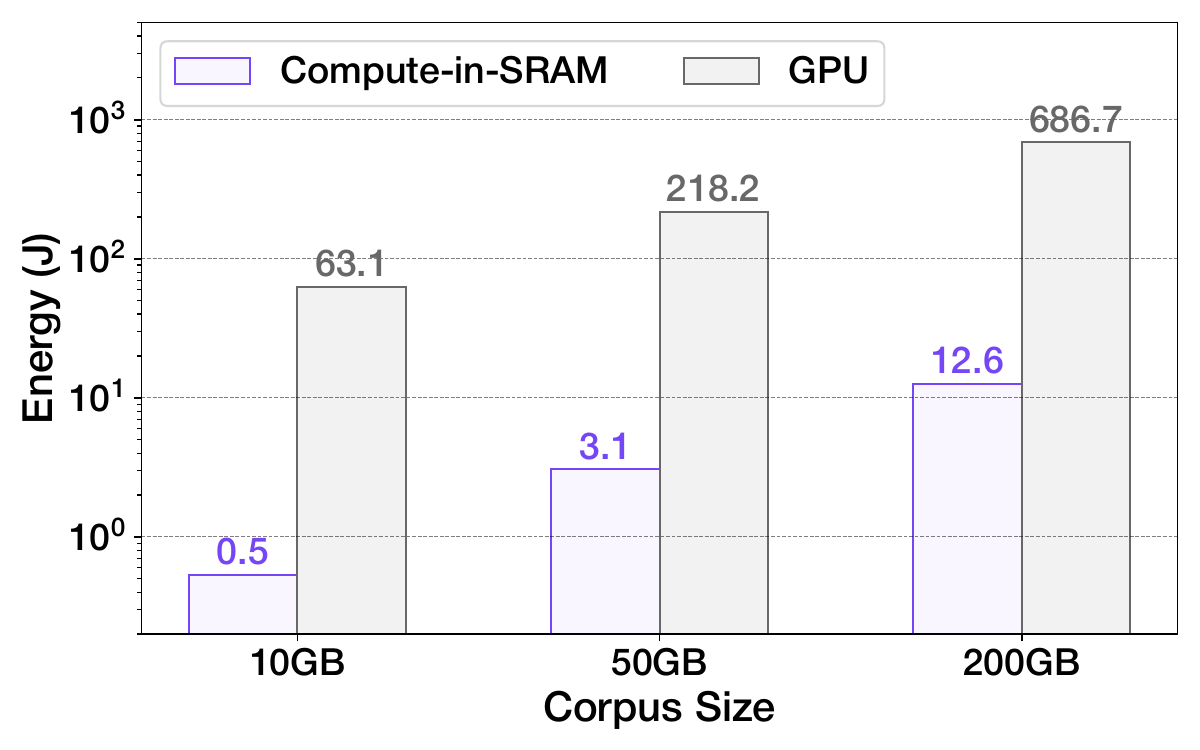}
    \vspace{-.1in}
    \caption{\revise{Top-5 retrieval process energy comparison with GPU. Results are measured on GSI Leda-E APU and NVIDIA A6000 GPU.}}
    \vspace{-.2in}
    \label{fig:rag-energy}
\end{figure}

We benchmark top-5 retrieval on our optimized compute-in-SRAM accelerator against an NVIDIA A6000 GPU, measuring GPU energy with \texttt{nvidia-smi}. As shown in Fig.~\ref{fig:rag-energy}, the APU is 54.4$\times$–117.9$\times$ more energy-efficient than the GPU. At 200~GB, APU energy is dominated by static (71.4\%), followed by compute (24.7\%), DRAM (2.7\%), other (1.1\%), and cache (0.005\%); smaller corpora show similar distributions, indicating static power dominates while compute scales modestly with workload size.

\input{tables/rag-breakdown}

%% file: tables/phoenix.tex
\begin{table}[t]
\centering
\caption{Statistics of the phoenix benchmark suite.}
\vspace{-.1in}
\resizebox{\linewidth}{!}{
\begin{tabular}{>{\itshape}lccc}
\toprule
\textbf{Application} & \textbf{Input Size} & \textbf{\#Inst. on CPU} & \textbf{\#Inst. of APU $\mu$Code} \\ \midrule
Histogram            & 1.5GB               & 4.8 billion                & 110.7 million    \\ 
Linear Regression    & 512MB               & 3.8 billion                & 1.6 million      \\
Matrix Multiply      & 1,024$\times$1,024  & 22.6 billion               & 69.7 million     \\ 
Kmeans               & 128k                & 0.4 billion                & 0.04 million     \\ 
Reverse Index        & 100MB               & 4.8 billion                & 11.0 million     \\
String Match         & 512MB               & 101.8 billion              & 0.09 million     \\ 
Word Count           & 10MB                & 0.7 billion                & 0.17 million     \\ \bottomrule
\end{tabular}
\label{tab:phoenix}
}
\vspace{-.1in}
\end{table}

%% file: tables/validation.tex
\begin{table}[t]
\centering
\caption{Phoenix benchmark suite latency measured vs. analytical framework.}
\vspace{-.1in}
\resizebox{.9\linewidth}{!}{
\begin{tabular}{>{\itshape}lccc}
\toprule
\textbf{Application} & \textbf{Meas. Latency (ms)} & \textbf{Predicted (ms)} & \textbf{Error}\\ \midrule
Histogram            & 1644.8             & 1650.1          & +0.32\%   \\ 
Linear Regression    & 92.3               & 94.5            & +2.3\%    \\
Matrix Multiply      & 421.3              & 402.5           & -4.5\%    \\ 
Kmeans               & 1.6                & 1.4             & -6.2\%    \\ 
Reverse Index        & 182.0              & 181.1           & -0.49\%    \\
String Match         & 90.9               & 92.6            & +1.8\%     \\ 
Word Count           & 3.2                & 3.1             &  -3.1\%    \\ \bottomrule
\end{tabular}
\label{tab:validation}
}
\vspace{-.1in}
\end{table}

%% file: tables/rag-breakdown.tex
\begin{table}[t]
\centering
\caption{\revise{Compute-in-SRAM retrieval latency breakdown across corpus size with and without optimizations.}}
\resizebox{\linewidth}{!}{
\begin{tabular}{lcccccc}
\hline
\textbf{} & \multicolumn{3}{c}{\textbf{Compute-in-SRAM No Opt}} & \multicolumn{3}{c}{\textbf{Compute-in-SRAM All Opts}} \\\cline{2-7}
\textbf{Corpus Size} & 10 GB & 50 GB & 200 GB & 10 GB & 50 GB & 200 GB \\
\hline
Load Embedding* & 0.4 ms & 2.0 ms & 8.2 ms & 0.3 ms & 1.5 ms & 6.1 ms \\
Load Query & 10 µs & 11 µs & 10 µs & 62 µs & 62 µs & 65 µs \\
Calc Distance & 21.0 ms & 126.5 ms & 527.9 ms & 3.1 ms & 18.0 ms & 74.6 ms \\
Top-K Aggregation & 69 µs & 325 µs & 1.30 ms & 72 µs & 317 µs & 1.24 ms \\
Return Top-K & 14 µs & 14 µs & 14 µs & 15 µs & 16 µs & 16 µs \\ \hline
\textbf{Total} & \textbf{21.8 ms} & \textbf{129.5 ms} & \textbf{539.2 ms} & \textbf{3.9 ms} & \textbf{20.6 ms} & \textbf{84.2 ms} \\
\hline
\end{tabular}
}
\label{tab:rag-breakdown}
  \raggedright\footnotesize
\revise{* Load embedding latency reflects simulated HBM2e performance; all other values are measured on GSI APU hardware.}
\vspace{-.2in}
\end{table}

%% file: sections/6-related-work.tex
\section{Related Work}

\noindent \textbf{Compute-in-Memory} architectures \revise{hold} the promise of being a highly energy-efficient approach for data-intensive applications by reducing data movement between memory and compute units. The Intelligent RAM (IRAM)~\cite{patterson1997intelligent, patterson1997case, oskin1998active} was one of the earliest efforts to integrate computational logic directly into DRAM, demonstrating the potential of coupling memory with vector processing. Building on this idea, VIRAM~\cite{gebis2004viram1} introduced a full vector processor with embedded DRAM to accelerate bandwidth-bound workloads. DIVA~\cite{ahn2015scalable} brought SPMD (single-program multiple-data) models to Processing-in-Memory (PNM), enabling more flexible parallelism, while FlexRAM~\cite{ahn2015pim, brockman2004low} extended this model within embedded DRAM systems, highlighting the importance of programmable abstractions for general-purpose compute-in-memory platforms.

\noindent \textbf{Compute-in-SRAM} architectures has been explored to realize boolean~\cite{agrawal2018x, agrawal2019xcel, bankman2018always, jeloka201628}, multiply-and-accumulate (MAC)~\cite{al2020towards, biswas2018conv, imani2019floatpim, ishida1998novel, lee2021charge, zhang2022pimca, laguna2021memory, laguna2022hardware, tu2022trancim}, and associative computing~\cite{potter1994asc, sayre1976staran, foster1976content, guo2013ac} mechanisms. Jeloka et al.~\cite{jeloka201628} introduced bit-line compute techniques in SRAMs, enabling bitwise logical operations between rows. Compute caches~\cite{aga2017compute} applied bit-line compute to transform chip multiprocessor (CMP) caches into logical compute engines. SRAM-based technologies have also proven effective for in-situ MAC operations due to the high on/off impedance ratio of SRAM bit cells~\cite{biswas2018conv, lee2021charge, jiang2020c3sram, zhang2017memory}. Associative computing, which uses primitives like search and multi-write to achieve in-memory compute~\cite{foster1976content, sayre1976staran}, has seen renewed interest with modern technologies. CAPE~\cite{caminal2021cape}, for example, demonstrates a CMOS-based associative engine with high programmability and low area cost.

\noindent \textbf{APU Microbenchmarking.} Prior work mapped RISC-V vector abstractions to the APU~\cite{golden2023supporting}, accelerated genomics kernels~\cite{golden2024accelerating}, and implemented cryptographic primitives~\cite{lee2023evaluating}, but these hand\-tuned microkernels highlight architectural features rather than system-level behavior. In contrast, we evaluate end-to-end Phoenix and RAG workloads, providing detailed performance characterization, and analysis of realistic compute and memory demands.

%% file: sections/7-conclusion.tex
\section{Conclusion}

This work provides a comprehensive evaluation of compute-in-SRAM devices under realistic workloads. Our analytical framework highlights key optimizations for general-purpose in-SRAM computing. With communication-aware reduction mapping, coalesced DMA, and broadcast-friendly data layouts, we accelerated RAG retrieval stage by 4.8$\times$--6.6$\times$ and reduced time-to-interactive latency by 1.1$\times$--1.8$\times$ over an optimized CPU baseline. Our system matched the performance of an NVIDIA A6000 GPU while consuming 54.4$\times$--117.9$\times$ less energy, underscoring the practicality and efficiency of compute-in-SRAM architectures.

%% file: sections/8-acknowledgements.tex
\textbf{Acknowledgments} --
This work was supported in part by the NSF PPoSS Award \#2118709, the NSF Graduate Research Fellowship Program (GRFP) Award \#2141064, and the ACE Center for Evolvable Computing, one of the seven centers in JUMP 2.0.